\documentclass{article}

\usepackage{microtype}
\usepackage{graphicx}
\usepackage{booktabs} 
\usepackage{subfig}     

\usepackage{hyperref}
\usepackage{amsthm}
\usepackage{comment}
\usepackage{pifont}

\usepackage[accepted]{icml2025}

\usepackage{amsmath}
\usepackage{amssymb}
\usepackage{mathtools}
\usepackage{amsthm}
\usepackage{algorithm}
\usepackage{algorithmic}
\usepackage{xcolor}
\definecolor{royalblue}{RGB}{65, 105, 225}

\usepackage[capitalize,noabbrev]{cleveref}

\theoremstyle{plain}

\theoremstyle{definition}

\theoremstyle{remark}

\usepackage[textsize=tiny]{todonotes}
\renewcommand{\algorithmiccomment}[1]{\hfill \textit{// #1}}

\definecolor{DarkGreen}{rgb}{0.0, 0.5, 0.0} 

\icmltitlerunning{LotteryCodec: Searching the Implicit Representation in a Random Network for Low-Complexity Image Compression}

\begin{document}

\twocolumn[


\icmltitle{LotteryCodec: Searching the Implicit Representation in a Random Network  \\ for Low-Complexity Image Compression}



\icmlsetsymbol{equal}{*}

\begin{icmlauthorlist}
\icmlauthor{Haotian Wu}{yyy}
\icmlauthor{Gongpu Chen}{yyy}
\icmlauthor{Pier Luigi Dragotti}{yyy}
\icmlauthor{Deniz Gündüz}{yyy}

\end{icmlauthorlist}

\icmlaffiliation{yyy}{Department of Electrical and Electronic Engineering, Imperial College London, London SW7 2AZ, U.K.}
\icmlcorrespondingauthor{Gongpu Chen}{gongpu.chen@imperial.ac.uk}
\icmlcorrespondingauthor{Haotian Wu}{haotian.wu17@imperial.ac.uk}

\icmlkeywords{Machine Learning, ICML}

\vskip 0.3in
]



\printAffiliationsAndNotice{} 

\begin{abstract}
We introduce and validate the \textit{lottery codec hypothesis}, which states that untrained subnetworks within randomly initialized networks can serve as synthesis networks for overfitted image compression, achieving rate-distortion (RD) performance comparable to trained networks. This hypothesis leads to a new paradigm for image compression by encoding image statistics into the network substructure. Building on this hypothesis, we propose LotteryCodec, which overfits a binary mask to an individual image, leveraging an over-parameterized and randomly initialized network shared by the encoder and the decoder. To address over-parameterization challenges and streamline subnetwork search, we develop a rewind modulation mechanism that improves the RD performance. LotteryCodec outperforms VTM and sets a new state-of-the-art in single-image compression. LotteryCodec also enables adaptive decoding complexity through adjustable mask ratios, offering flexible compression solutions for diverse device constraints and application requirements. Project page: \url{https://eedavidwu.github.io/LotteryCodec/}
\end{abstract}

\section{Introduction}
\label{intro}
Traditional image/video compression algorithms rely on meticulously designed linear transforms. Recently, conventional methods have been increasingly challenged by emerging learning-based codecs that replace analysis and synthesis transforms in classical codecs with neural networks~\cite{balle2018variational, cheng2020image, he2022elic}. While achieving impressive performance gains, these autoencoder (AE)-based neural codecs often suffer from high decoding complexity and large number of network parameters, which limit their practical deployment on resource-constrained devices \cite{jiang2023mlic, wang2023evc}. In addition, they require training on extremely large datasets to ensure robust performance across diverse image distributions --- resources that are not always available. Addressing these issues to develop low-complexity and robust codecs with competitive rate-distortion (RD) performance remains a critical open challenge~\cite{yang2023introduction}.

\begin{figure}[t]
    \centering
\includegraphics[width=1\linewidth,page=1]{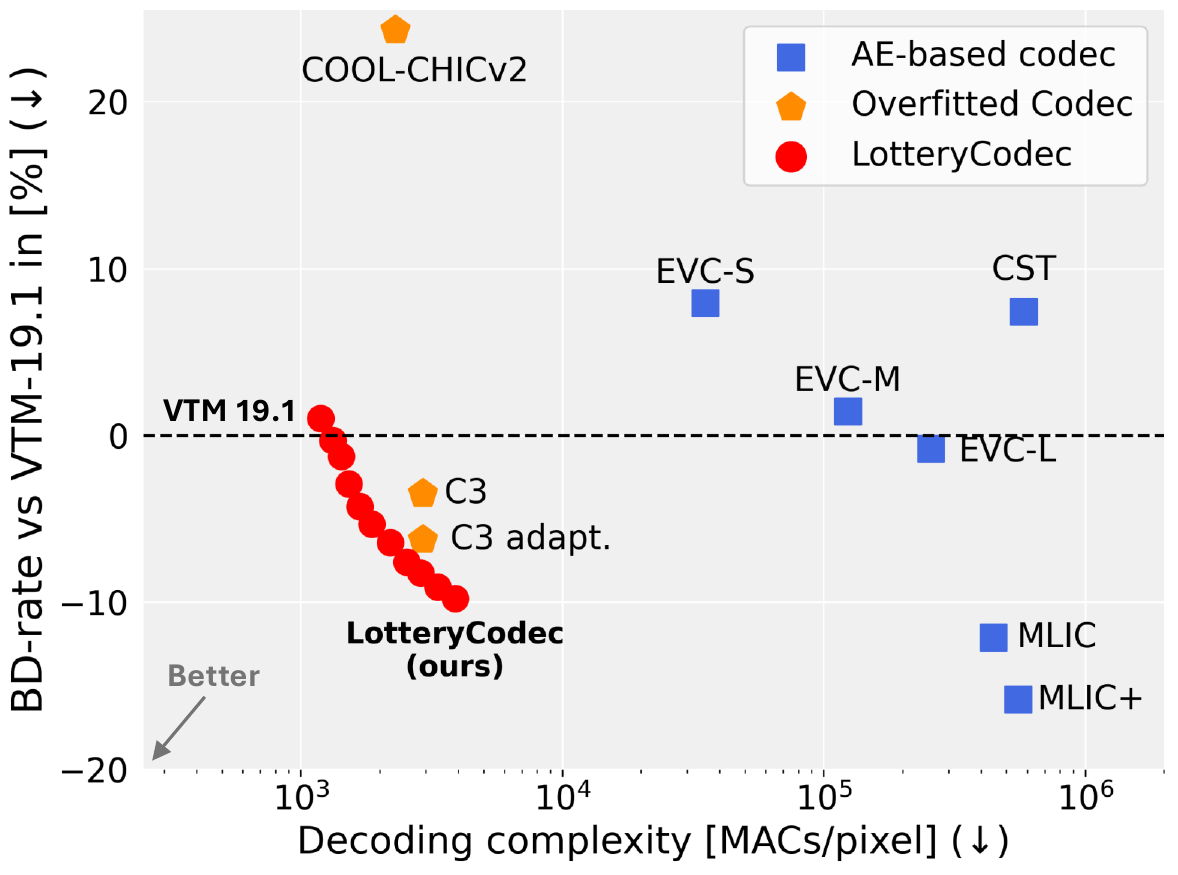}
    \caption{Rate-distortion performance (BD-rate) vs. decoding complexity on the CLIC2020 dataset. LotteryCodec achieves a superior and adaptable RD trade-off than other codecs.}
    \label{fig:complexity_clic}
\end{figure}

Implicit neural representations (INRs) \cite{sitzmann2020implicit} has emerged as a promising signal representation technique 
that leverages lightweight multi-layer perceptrons (MLPs) to directly parameterize continuous functions. INRs have demonstrated impressive performance in various modalities and tasks, including modality-agnostic representation~\cite{shi2024improved}, 3D reconstruction~\cite{atzmon2020sal}, and view synthesis~\cite{mildenhall2021nerf}, and have been extended to multi-instance settings via modulation mechanisms \cite{mehta2021modulated}.
Quantizing the overfitted network yields a signal compressor. For image compression, a lightweight neural network can overfit a single image by mapping pixel coordinates to their corresponding intensities. 
Building on this idea, the first INR-based overfitted image codec, COIN, was proposed in \cite{dupont2021coin}. Later, the COOL-CHIC series codecs \cite{ladune2023cool,leguay2023low} and C3 \cite{kim2024c3} further enhanced the RD performance, outperforming widely used codecs such as BPG \cite{bellard2015bpg}, HEVC \cite{sullivan2012overview}, and VCC \cite{bross2021developments}, while maintaining low decoding complexity.
Despite significant advances, state-of-the-art overfitted image codecs still fall short of the RD performance of competitive classical codecs such as VTM~\cite{bross2021overview}. Further improvements are challenging because achieving higher reconstruction fidelity typically necessitates larger networks, which significantly increases the compression rate. 
In addition, most overfitted codecs rely on fixed network architectures for a range of images, limiting their RD performance and adaptability. 
We thus anticipate a new paradigm, one that can balance the increasing cost of network complexity with representation capability while enabling an adaptive architecture for individual images. 

This work is inspired by two key findings: (1) a randomly initialized neural network can act as a handcrafted prior, encoding significant image statistics and prior information within its network structure \cite{ulyanov2018deep}; and (2) over-parameterized neural networks contain high-performing untrained subnetworks~\cite{ramanujan2020s}. These insights motivate us to propose the {\textit{lottery codec hypothesis}}. Specifically, consider a well-trained overfitted image codec $g_{\mathbf{W}}(\mathbf{z})$, which represents image $\mathbf{S}$ as a neural network parameterized by $\mathbf{W}$, with a latent representation $\mathbf{z}$ as its input. After quantization and entropy coding, the decoder reconstructs the image as $\mathbf{{S}}^* = g_{\hat{\mathbf{W}}}(\mathbf{\hat{z}})$ \footnote{In this paper, variables with a hat notation, such as $\mathbf{\hat{z}}$, represent their quantized counterparts.}. Next, consider an over-parameterized, randomly initialized network $g_{\mathbf{W'}}$, along with a learned binary mask $\mathbf{\tau'}\in \{0,1\}^{|\mathbf{W'}|}$ and a latent $\mathbf{z'}$, where $|\mathbf{W'}|$ denotes the number of parameters of $g_{\mathbf{W'}}$. The source image is reconstructed via a subnetwork of $g_{\mathbf{W'}}$ as $\mathbf{S}'=g_{\mathbf{W'}\odot \mathbf{\tau'}}(\mathbf{\hat{z}'})$, where $\odot$ represents the Hadamard product, identifying the subnetwork. We propose the following hypothesis:

\noindent\textbf{Lottery codec hypothesis.} 
\textit{ Let $d$ denote a distortion function and $H$ the entropy function. For any overfitted image codec $g_{\mathbf{W}}(\mathbf{z})$, there exists a pair $(\mathbf{\tau'},\mathbf{z'})$ as the ‘winning tickets’ within a sufficiently over-parameterized and randomly initialized network $g_\mathbf{W'}$ satisfying $|\mathbf{W'}|>|\mathbf{W}|$, such that $d(\mathbf{S},\mathbf{S}') \le d(\mathbf{S},\mathbf{{S}}^*)$ and $H(\mathbf{\hat{z}}')=H(\mathbf{\hat{z}})$. }

Note that the equality $H(\mathbf{\hat{z}}’) = H(\mathbf{\hat{z}})$ implies that the number of bits required to represent $\mathbf{\hat{z}}’$ and $\mathbf{\hat{z}}$ after entropy coding is the same.  We conjecture this hypothesis to hold for the following reasons: (1) Theoretical justification: prior studies \cite{pensia2020optimal,da2022proving} suggest that any target network of width $l_w$ and depth $l_d$ can be approximated by pruning a random network that is a factor $O(\log(l_wl_d))$ wider and twice as deep, suggesting that sufficiently over-parameterized, randomly initialized networks contain some ‘winning tickets’ even without training. (2) Empirical evidence: extensive experiments conducted in this paper evaluate the hypothesis, and the results consistently support its validity.
\begin{figure}[t]
    \centering
\includegraphics[width=1\linewidth,page=1]{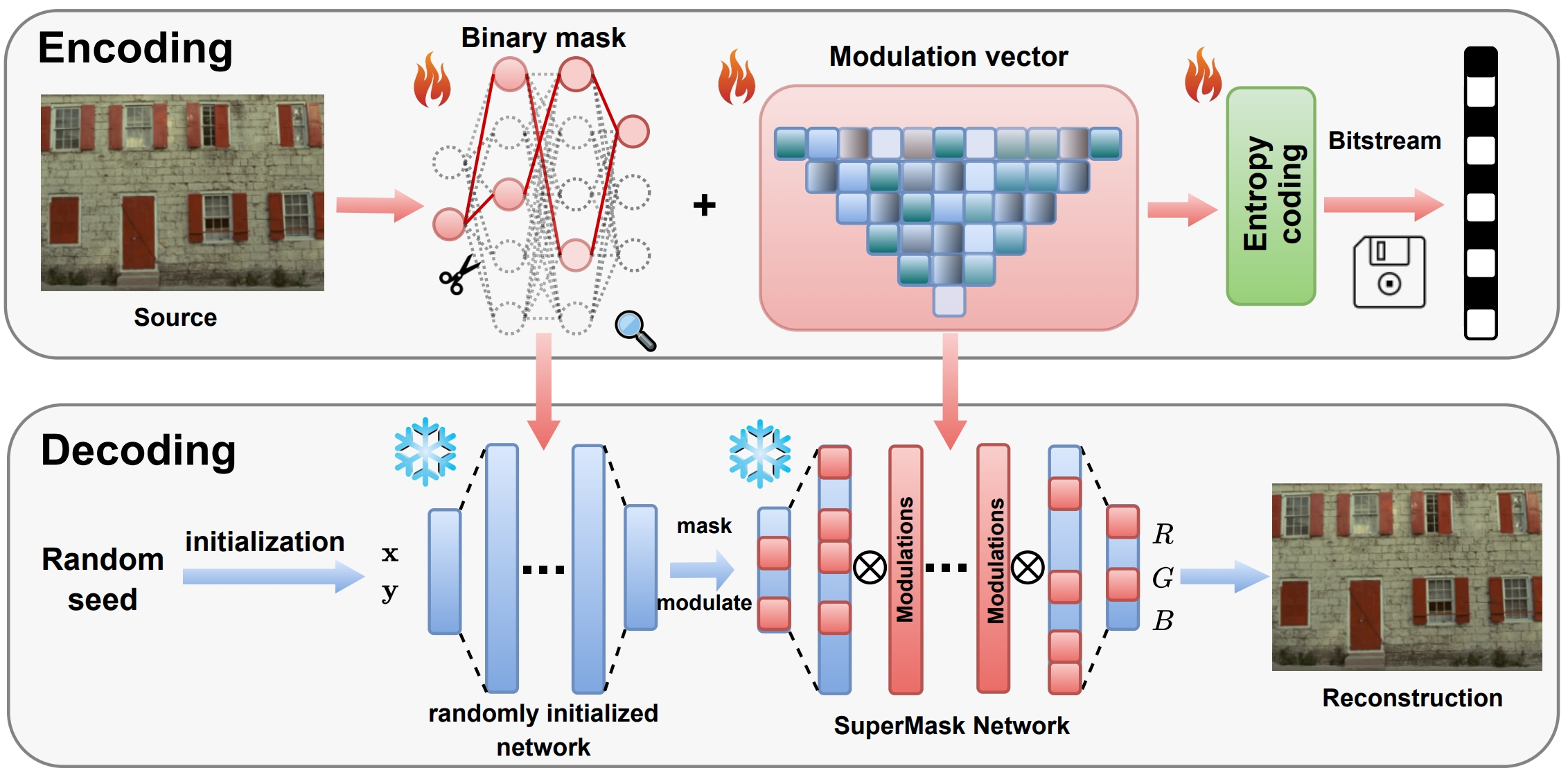}
    \caption{Illustration of LotteryCodec scheme: the source image is encoded into a binary mask and latent modulations. During decoding, the receiver initializes a common random network and uses a modulated subnetwork to reconstruct the source image.
    }
    \label{fig:figure2}
\end{figure}

The proposed \textit{lottery codec hypothesis} highlights the potential of searching untrained but well-performing subnetworks as overfitted image codecs. While theoretical works and experiments demonstrate that sufficiently over-parameterized networks can contain untrained subnetworks that match the performance of well-trained networks, precise guidelines on the required level of over-parameterization, such as specific architectural depth or configurations, remain unclear for reliably obtaining a ‘winning ticket’.

\begin{figure*}[t]
    \centering
    \includegraphics[width=1\linewidth]{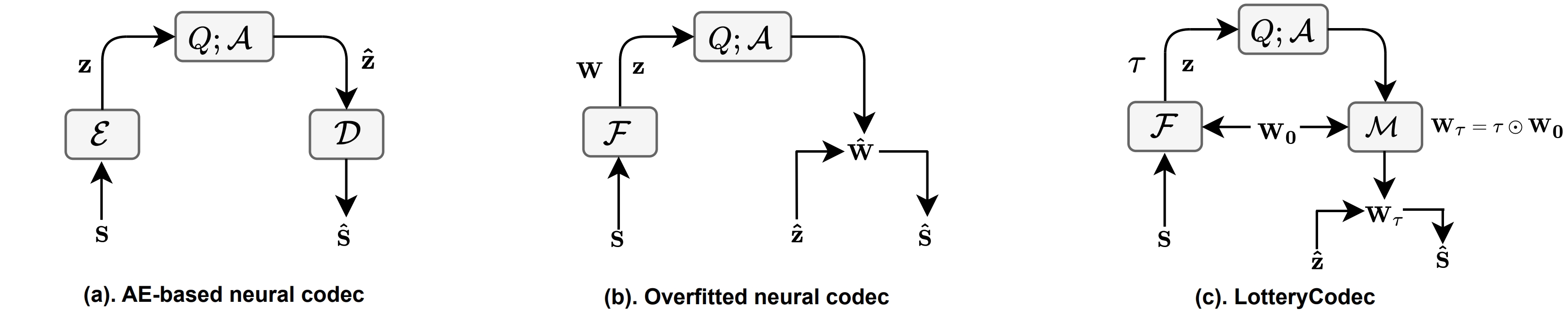}
    \caption{Operational structure of different compression schemes. (a) AE-based neural codecs: source image $\mathbf{S}$ is processed through a pair of  encoder and decoder. (b) Overfitted neural codecs: $\mathbf{S}$ is fitted by parameters $\{\mathbf{W},\mathbf{z}\}$ via a fitting operation $\mathcal{F}$. 
    (c) LotteryCodec: $\mathbf{S}$ is fitted by parameters $\{\mathbf{\tau},\mathbf{z}\}$, identifying a subnetwork in a randomly initialized network, with masking operations $\mathcal{M}$. } 
    \label{fig:compress_model}
\end{figure*}

Based on the above hypothesis, we propose a novel overfitted image compression scheme, called LotteryCodec (see Fig. \ref{fig:figure2}). This method utilizes a randomly initialized network as the synthesis network and learns a binary mask with latent modulations as the code. With a predefined random seed, network parameters are eliminated from transmissions; instead, it is sufficient to transmit only the binary mask and latent variables for the decoder to reconstruct the image. 

To address the challenges of over-parameterization and subnetwork search, we propose a rewind modulation mechanism that introduces image information during the subnetwork search, simplifying the process and enhancing the RD performance. 
\textit{To the best of our knowledge, the proposed LotteryCodec is the first overfitted image codec to surpass the RD performance of VTM while maintaining a low decoding complexity.} Furthermore, LotteryCodec sets a new state-of-the-art for overfitted neural codecs obtained from a single image. Its adaptive masking strategy enables flexible model complexity adjustment based on varying mask ratios, balancing computational cost and performance, as shown in Fig. \ref{fig:complexity_clic}.


The contributions of this work are summarized as follows:
\begin{itemize}
    \item We propose and experimentally verify the \textit{lottery codec hypothesis}, which suggests that a subnetwork within a randomly initialized neural network can directly serve as a well-performing synthesis network for overfitted image compression. This hypothesis introduces a new paradigm for INR-based image compression, emphasizing the potential of encoding image statistics into the structure of a randomly initialized network.

    \item We propose a novel LotteryCodec scheme, which overfits a randomly initialized neural network to the source image by learning a binary mask and modulation vectors. To alleviate over-parameterization and simplify subnetwork search, a rewind modulation mechanism is introduced, significantly improving the RD performance.   
    
    \item We show by extensive experiments that LotteryCodec achieves state-of-the-art performance among overfitted image codecs designed for single-image compression at a reduced computational cost. Additionally, LotteryCodec can adjust its decoding complexity by varying the mask ratio, thus providing flexible solutions for diverse computational and performance needs.
    
\end{itemize}

\section{Related work}
\subsection{Neural data compression}
Currently, there are two main paradigms for neural data compression (see Fig. \ref{fig:compress_model}): AE-based and overfitted neural codecs, both designed to balance the trade-off between the distortion and the rate~\cite{cover1999elements}.

\noindent\textbf{AE-based neural codecs.} As illustrated in Fig. \ref{fig:compress_model}a, an AE-based neural codec, e.g., \cite{balle2018variational,cheng2020image,he2022elic,jiang2023mlic,wang2023evc}, comprises a pair of encoding network $\mathcal{E}$ and decoding network $\mathcal{D}$, jointly optimized over a large dataset. A source image $\mathbf{S}$ is encoded by $\mathcal{E}$ into a latent representation $\mathbf{z}$, which is then compressed into $R$ bits by quantization and entropy coding. At the decoder side, the quantized latent representation $\hat{\mathbf{z}}$ is first recovered and then used to reconstruct the source image. Resulting average distortion is $D = \mathbb{E}_{\mathbf{S} \sim p_s} \left[d(\mathbf{S},\mathcal{D}(\mathbf{\hat{z}}) )\right]$, where $d$ denotes the distortion metric, and the compression rate $R$ is given by:
\begin{align}
  R = \mathbb{E}_{\mathbf{S} \sim p_s} \left[-\log_2 p_{Q\left(\mathcal{E}(\mathbf{S})\right)} \left( {\mathbf{\hat{z}}}\right)\right],
\end{align}
where $Q $ represents quantization operations. Although AE-based approaches achieve competitive performance, their decoding complexity is typically high due to reliance on large and complex architectures for robust latent representations (see Fig. \ref{fig:complexity_clic}).


\noindent\textbf{Overfitted neural codecs.}
Overfitting a parametric function to each single image offers an alternative, where quantized function parameters serve as the compressed representation. A low-complexity network is often sufficient since generalization is not required. COIN~\cite{dupont2021coin} pioneered this by training a lightweight MLP to map pixel coordinates to RGB values. Quantized parameters are entropy-coded into a bitstream for decoding. COIN++~\cite{dupont2022coin++} improved generalization with meta-learning. However, COIN’s RD performance remains limited despite low complexity and remarkable generalization. Subsequently, the COMBINER series ~\cite{combiner,he2024recombiner} introduces variational INRs and leverages relative entropy coding for RD-optimized compression. 
\begin{figure*}[t]
    \centering
\includegraphics[width=1\linewidth]{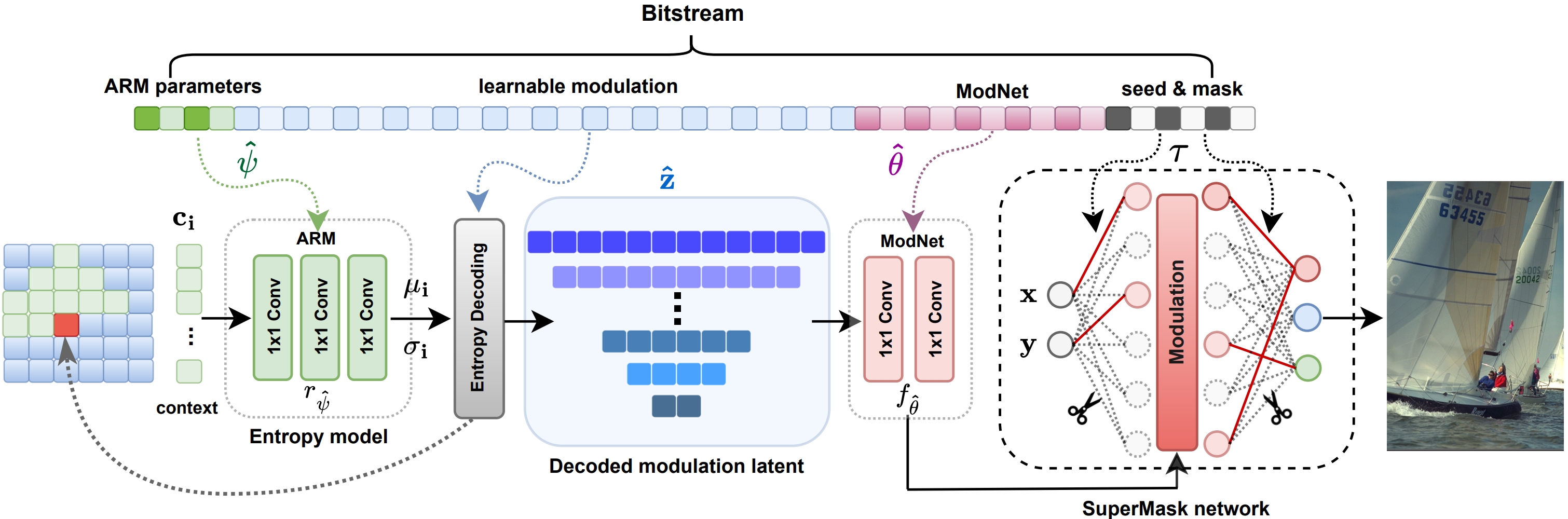}
    \caption{Illustration of the image decoding process in LotteryCodec. The ARM parameters $\mathbf{\hat{\psi}}$ are first retrieved to regress the latent modulations $\mathbf{\hat{z}}$. Subsequently, the binary mask $\mathbf{\tau}$ and initialization seed configure the synthesis network, while the modulation model parameters $\mathbf{\hat{\theta}}$ are decoded to generate the modulations from $\mathbf{\hat{z}}$ to guide the synthesis network to reconstruct the image.}
    \label{fig:decoding_fig}
\end{figure*}

A significant advancement came with COOL-CHIC ~\cite{ladune2023cool}, which improves RD performance by integrating a latent representation and an entropy model while maintaining a low decoding complexity. As illustrated in Fig. \ref{fig:compress_model}b, COOL-CHIC jointly trains a parametric function $g_{\mathbf{W}}$ with a latent vector $\mathbf{z}$ as the input to overfit $\mathbf{S}$. Similarly to COIN, both $\mathbf{W}$ and $\mathbf{z}$ are compressed into $R$ bits through quantization and entropy coding. During decoding, the quantized network and latent vector are used to reconstruct the image via the mapping $\mathbf{\hat{S}}=g_{\mathbf{\hat{W}}}(\mathbf{\hat{z}})$. This results in average distortion $D =\mathbb{E}_{\mathbf{S}\sim p_s}\left[d({\mathbf{S}},g_{\mathbf{\hat{W}}}(\mathbf{\hat{z}}))\right]$ at a rate 
\begin{align}
    R  = \mathbb{E}_{\mathbf{S} \sim p_s} \bigg[
    -\log_2 p_{\hat{\psi}} \big( {\mathbf{\hat{z}}}\big)
     -\log_2 p ( {\mathbf{\hat{W}}}) + R_{\hat{\psi}}  \bigg],
    \label{eq: RD-2}
\end{align}
where $\mathcal{F}$ denotes the operations that overfit $(\mathbf{W},\mathbf{z})$ parameters to the specific input $\mathbf{S}$, 
 $p_{\hat{\psi}}(\cdot)$ is the estimated distribution with estimation model $\hat{\psi}$. In practice, entropy coding of ${\mathbf{\hat{z}}}$ relies on a lightweight auto-regressive entropy model (ARM) $r_{\psi}$ for distribution estimation. In \eqref{eq: RD-2}, $R_{\hat{\psi}}$ represents the extra bit overhead due to the transmission of this model parameters. 

Extensions like COOL-CHICv2 \cite{leguay2023low}, C3 \cite{kim2024c3,balle2024good}, and COOL-CHICv3 \cite{blard2024overfitted} further enhance the RD performance with advanced architectures and techniques, including soft-rounding, Kumaraswamy noise, and conditional entropy models. These advancements allow COOL-CHIC to outperform widely used codecs such as BPG and HEVC. 

\subsection{Lottery ticket hypothesis}
\citet{frankle2018lottery} introduced the \textit{lottery ticket hypothesis} (LTH), stating that a randomly initialized, dense neural network contains a subnetwork that can match the test accuracy of the full network with equivalent training. \citet{zhou2019deconstructing} found that ‘winning tickets’ (i.e., masked subnetworks) can outperform random initialization without training, while \citet{ramanujan2020s} extended the idea by showing that even untrained subnetworks can achieve near state-of-the-art performance, a phenomenon referred to as the \textit{strong lottery ticket hypothesis} (SLTH). Building on these ideas, \citet{choi2023overfitting} applied LTH to video representation with multiple supermask overlays and unpruned biases. This LTH-based video representation method enhances expressiveness but increases complexity.

Recently, \citet{malach2020proving} theoretically proved the SLTH for fully connected networks with ReLU activations, showing that any target network can be approximated by pruning a sufficiently over-parameterized random network of polynomial size relative to the target network. \citet{orseau2020logarithmic} and \citet{pensia2020optimal} relaxed the assumptions and improved these bounds to logarithmic order, which is later extended into convolutional neural networks (CNNs) in \cite{da2022proving}.

\section{LotteryCodec}
\label{sec:method}
Inspired by the \textit{lottery codec hypothesis}, we propose LotteryCodec, a novel image compression paradigm that encodes images by identifying structured subnetworks within a randomly initialized network. The detailed pseudocode for the method is provided {in Appendix \ref{pse_algorithm}}.

\subsection{Workflow}


\noindent\textbf{Encoder.} 
As depicted in Fig \ref{fig:compress_model}(c), given an over-parameterized network $g_{\mathbf{W_0}}$ with randomly initialized parameters $\mathbf{W}_0$, each image $\mathbf{S}$ is overfitted using a subnetwork of $g_{\mathbf{W_0}} $ by learning a binary mask $\mathbf{\tau}$ and a latent representation $\mathbf{z}$, such that the subnetwork $g_{\mathbf{\tau}\odot \mathbf{W_0}}$ can well reconstruct $\mathbf{S}$ using $\mathbf{\hat{z}}$. The quantized latent $\mathbf{\hat{z}}$ and binary mask $\mathbf{\tau}$ are then compressed via entropy coding, producing a bitstream representation of $\mathbf{S}$. Additional mechanisms, such as ARM $r_\psi$ and the modulation model ModNet $f_\theta$, are incorporated to further enhance the RD performance, as will be elaborated on later.
\begin{figure*}[t]
    \centering
    \includegraphics[width=1\linewidth]{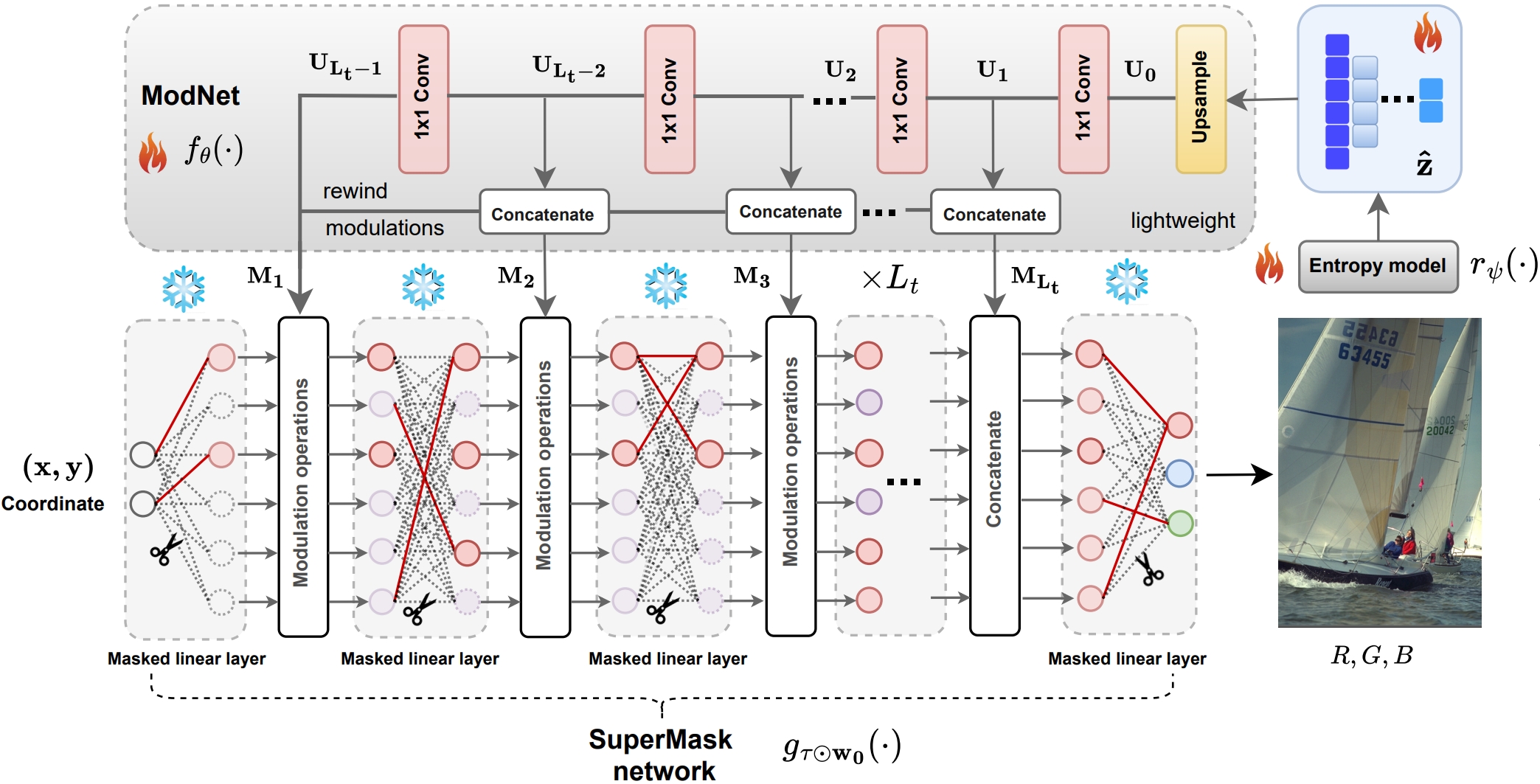}
    \caption{Illustration of the ModNet and SuperMask networks: SuperMask network maps pixel coordinates to RGB values by identifying subnetworks within a randomly initialized network, guided by modulations generated by the ModNet using input $\hat{\mathbf{z}}$. Solid red lines indicate active weights, while dashed lines represent masked weights. The weights of the SuperMask network remain frozen during training; only the binary mask $\mathbf{\tau}$ and ModNet parameters are learned.}
    \label{fig:mask}
\end{figure*}

\noindent\textbf{Decoder.} 
As shown in Fig.~\ref{fig:decoding_fig}, 
the decoder begins by initializing the neural network $g_{\mathbf{W}0}$ using the same random seed as the encoder. Next, the binary mask $\tau$ is extracted to identify a subnetwork $g_{\mathbf{\tau}\odot \mathbf{w_0}}$, and the ARM parameters $\mathbf{\hat{\psi}}$ are retrieved to decode the latent modulation vector $\mathbf{\hat{z}}$.
Finally, a modulation model $f_{\mathbf{\hat{\theta}}} $ is retrieved and applied to modulate the image generation process based on $\mathbf{\hat{z}}$ as:
\begin{equation}
    \mathbf{\hat{S}}(\mathbf{x})=g_{\mathbf{\tau}\odot \mathbf{w_0}}(f_{\mathbf{\hat{\theta}}} (\mathbf{\hat{z}}),\mathbf{x}),
\end{equation}
where $\mathbf{x}$ denotes the pixel coordinate vector of the image. The distortion of LotteryCodec is then measured by $D  = \mathbb{E}_{\mathbf{S} \sim p_s} \left[ d(\mathbf{S}, \mathbf{\hat{S}})\right]$. 
The decoding complexity remains low because $g_{\mathbf{\tau}\odot \mathbf{w_0}}$ is a lightweight network. Moreover, LotteryCodec offers flexible decoding complexity by allowing adjustable mask ratios, which dynamically control the active size of the synthesis network.

\noindent\textbf{RD cost optimization.} 
To balance the rate-distortion trade-off, LotteryCodec is trained to fit a parameter set, denoted by $\mathbf{\Omega}\triangleq\{\mathbf{{z}},\mathbf{{\psi}},\mathbf{{{\theta}}},\mathbf{{\tau}}\}$, with the goal of minimizing the following loss function (i.e., the RD cost):
\begin{equation}
    \mathcal{L}({\mathbf{\Omega}})= d(\mathbf{S},g_{\mathbf{\tau}\odot \mathbf{w_0}}(f_{\mathbf{\hat{\theta}}}(\mathbf{\hat{z}}),\mathbf{x}))+\lambda R(\mathbf{\hat{z}}),
    \label{r_d_eq_mask}
\end{equation}
where $R({\mathbf{\hat{z}}})$ represents the compression rate contributed by $\mathbf{\hat{z}}$, and $\lambda$ is a hyperparameter that controls the trade-off between distortion and rate. During training, $\mathbf{z}$ undergoes soft-rounding with varying Kumaraswamy noise, whereas hard-rounding is applied during inference. Notably, the loss function \eqref{r_d_eq_mask} excludes the rate terms associated with $\mathbf{\psi}$, $\mathbf{\theta}$, and $\mathbf{\tau}$, as their contribution to the overall bit rate is minimal due to their lightweight architectures. In practice, these parameters are quantized and entropy-coded using a non-learned distribution after the optimization process is completed, contributing to the final bitstream. 

Overall, the LotteryCodec bitstream consists of four parts: the ARM parameters $\mathbf{\hat{\psi}}$, learnable latent modulations $\mathbf{\hat{z}}$, modulation model parameters $\mathbf{\hat{\theta}}$, and the binary mask $\mathbf{\tau}$. Hence, the total compression rate is given by
\begin{align}
    R  = \mathbb{E}_{\mathbf{S} \sim p_s} \bigg[
    -\log_2 p_{\hat{\psi}}(\mathbf{\hat{z}}) -\log_2 p (\mathbf{\tau})+ R_{\hat{\theta}} + R_{\hat{\psi}} \bigg],
    \label{eq: RD-LTH}
\end{align}
where $R_{\hat{\theta}}$ and $R_{\hat{\psi}}$ denote the rate contributed by the transmission of quantized ModNet and ARM. As shown in Eqs. \eqref{eq: RD-2} and \eqref{eq: RD-LTH}, the rate of overfitted codecs depends on $\{\mathbf{\hat{z}}, \mathbf{\hat{\psi}}, \mathbf{\hat{W}}\}$, while the rate of our method is determined by $\{\mathbf{\hat{z}}, \mathbf{\hat{\psi}}, \mathbf{\tau}, \mathbf{\hat{\theta}}\}$. According to the Lottery Codec Hypothesis, our bit cost for $\mathbf{\hat{z}}$ and $\mathbf{\hat{\psi}}$ matches that of standard overfitted codecs. While each quantized parameter in $\hat{W}$ typically requires more than $13$ bits, our binary mask $\mathbf{\tau}$ uses up to $1$ bits per entry. Despite its higher dimensionality, $\tau$ contributes significantly less to the total rate. Moreover, since $\mathbf{\hat{\theta}}$ is lightweight, the combined rate of $\mathbf{\tau}$ and $\mathbf{\hat{\theta}}$ remains lower than that of $\mathbf{\hat{W}}$, resulting in an improved compression efficiency.
Note also that $\tau$ is transmitted in a lossless fashion as it does not need to be quantized. In practice, both $\tau$ and $\hat{\theta}$ are entropy-coded using offline-trained models or a static distribution.

\subsection{Winning a lottery codec}
The design of LotteryCodec incorporates four key architectural components: SuperMask network ($g_{\mathbf{W_0\odot \tau}})$, ModNet ($f_{\mathbf{\theta}}$), latent modulation ($\mathbf{z}$), and ARM ($r_{\mathbf{\psi}}$). As shown in Fig. \ref{fig:mask}, the SuperMask network $g_{\mathbf{\tau\odot\mathbf{W}_0}}$ is a high-performing subnetwork that maps pixel coordinates to RGB values, obtained by applying the learned mask $\tau$ to the over-parameterized network $g_{\mathbf{\mathbf{W}_0}}$. Specifically, $g_{\mathbf{\tau\odot\mathbf{W}_0}}$ comprises $L_t$ masked linear layers, with each masked linear layer followed by modulation operations. The ModNet $f_{\mathbf{\theta}}$ includes an upsampling operation and $(L_{t}-1)$ convolutional layers with $1\times1$ kernel to generate hierarchical modulation vectors for each layer of $g_{\mathbf{\tau\odot\mathbf{W}_0}}$ from the retrieved $\hat{\mathbf{z}}$. This design significantly simplifies the subnetwork search while improving the RD performance. The latent modulation $\mathbf{z}$ is a set of learnable vectors that serve as a primary compression component, while the ARM ($r_{\mathbf{\psi}}$) is an auto-regressive entropy model commonly used to compress latent representation $\mathbf{z}$ in overfitted codecs \cite{ladune2023cool}, with architecture detailed in Appendix \ref{quantization_entropy_coding}.

\paragraph{Fourier initialization.} Initialization plays a critical role in the SLTH problem to ensure the efficient identification of ``winning tickets'' \cite{ramanujan2020s}. In our LotteryCodec, parameters ${\mathbf{W_0}}$ are initialized using a Fourier initialization approach. This design is based on two main considerations: (1) From an INR perspective, the Fourier reparametrization method can address the low-frequency bias of MLPs and enhance INR with richer textures \cite{shi2024improved}; (2) From an SLTH point of view, this initialization ensures the network retains rich sign information \cite{zhou2019deconstructing,anonymous2024find} while maintaining constant variance between inputs and outputs \cite{glorot2010understanding,he2016deep}.

Specifically, the weight matrix of the $i$-th MLP layer, denoted by $\mathbf{W^{(i)}}\in \mathbb{R}^{d_i\times d_{i-1}}$, is re-parameterized as a weighted combination of fixed Fourier bases:
\begin{equation}
  \mathbf{W^{(i)}}=\mathbf{\Lambda^{(i)}}\mathbf{B^{(i)}},
     \label{eq_FFN_init}
\end{equation}
where $\mathbf{\Lambda^{(i)}}\in \mathbb{R}^{d_{i}\times M}$ is the coefficient matrix, and $\mathbf{B^{(i)}}\in \mathbb{R}^{M \times d_{i-1}}$ represents a set of $M$ Fourier bases. Each element $b^{(i)}_{m,n}$ of $\mathbf{B^{(i)}}$ (the $m$-th row and $n$-column) is defined using a distinct frequency and phase as $b^{(i)}_{m,n}=\cos(w_ma_n^{(i)}+\varphi_m)$, where $\mathbf{a^{(i)}}={a_1^{(i)},\ldots,a_{d_{i-1}}^{(i)}}$ is a positional sequence uniformly sampled from $[-\pi,\pi]$, and $\mathbf{w}$ and $\mathbf{\varphi}$ are the frequency and phase vectors. We adopt $P$ different phases and $2F$ different frequencies, hence $M=2FP$. The phase vector is defined as $\varphi\triangleq\{0,2\pi/P,\ldots,2\pi(P-1)/P\}$. For each phase, the frequency vector is defined as $\mathbf{w}\triangleq\{\mathbf{w_{low}},\mathbf{w_{high}}\}$, where $\mathbf{w_{low}}=\{1/F,2/F,\ldots,1\}$ and $\mathbf{w_{high}}=\{1,2,\ldots,F,\}$ denote low-frequency and high-frequency bases, respectively. Each element $\lambda_{m,n}^{(i)}$ of $\mathbf{\Lambda^{(i)}}$ is sampled from: 
\begin{equation*}
    \lambda_{m,n}^{(i)}\sim U\left(-\sqrt{\frac{6}{M\sum_{t=1}^{d_{i-1}}(b_{m,t}^{(i)})^2}},\sqrt{\frac{6}{M\sum_{t=1}^{d_{i-1}}(b_{m,t}^{(i)})^2}}\right).
\end{equation*}

\paragraph{SuperMask network.} We introduce a learnable matrix $\mathbf{P}$ to identify the subnetwork $g_{\tau\odot\mathbf{W}_0}$, defined by mask $\tau$, that minimizes the loss function. Matrix $\mathbf{P}$ shares the same dimensions as $\mathbf{W}_0$, with each element representing a score of the corresponding weight in $\mathbf{W_0}$. During training, the randomly initialized $\mathbf{W_0}$ remains frozen while only $\mathbf{P}$ is updated. The top $r_a\%$ of weights with the highest scores across all layers are activated, while the remaining weights are set to zero.

More formally, let $\mathbf{\mathcal{V}^{(i)}}\triangleq \{v_1^{(i)},\ldots,v_{d_{i}}^{(i)}\}$ denote the values of nodes for the $i$-th {masked linear layer} with $d_{i}$ nodes. The output of the $k$-th neuron in the $i$-th layer with the modulation function $m(\cdot)$ can then be written as:
\begin{equation}
    v_{k}^{(i)}=\sigma\left(\sum_{j=1}^{d_{i-1}}\tau_{k,j}^{(i-1)}w_{kj}^{(i-1)}m(v_{j}^{(i-1)}) \right),
    \label{mod_supermask}
\end{equation}
where $\sigma $ is the activation function and $w_{kj}^{(i-1)}$ is the weight connecting the $k$-th neuron in $i$-th layer to the $j$-th neuron of the $(i-1)$-th layer. The mask function returning the binary mask 
$\tau_{k,j}^{(i-1)}=h\left(\mathbf{P}\right)$
is defined as: $\tau_{k,j}^{(i-1)}=1$ if the score $p_{kj}^{(i-1)}$ for the weight is among the top $r_a\%$ highest values, and $\tau_{k,j}^{(i-1)}=0$ otherwise. Using the chain rule of gradient, the gradient of the loss $\mathcal{L}$ with respect to $p_{kj}^{(i-1)}$ can be estimated by a straight-through gradient estimator:
\begin{equation}  \label{eq: graident}
    \frac{\partial \mathcal{L}}{\partial p_{kj}^{(i-1)}} \approx \frac{\partial \mathcal{L}}{\partial v_{k}^{(i)}}\frac{\partial v_{k}^{(i)}}{\partial \sigma}w_{kj}^{(i-1)}m(v_{j}^{(i-1)}) \triangleq \varsigma^{(i-1)}_{kj}.
\end{equation}
The above formula holds because $\tau_{k,j}^{(i-1)}$ is a non-decreasing function of $p_{kj}^{(i-1)}$. The activation probability $p_{kj}^{(i-1)}$ can then be updated as:
\begin{equation}
    {p}_{kj}^{(i-1)}\leftarrow p_{kj}^{(i-1)}-\alpha \varsigma^{(i-1)}_{kj},
\end{equation}
where $\alpha$ is the learning rate. When activation or deactivation swaps occur, the loss decreases for the mini-batch, according to \cite{ramanujan2020s,wortsman2019discovering}. 


\begin{figure*}[t]
    \centering
    \subfloat[]{
        \centering
        \includegraphics[width=0.258\linewidth]{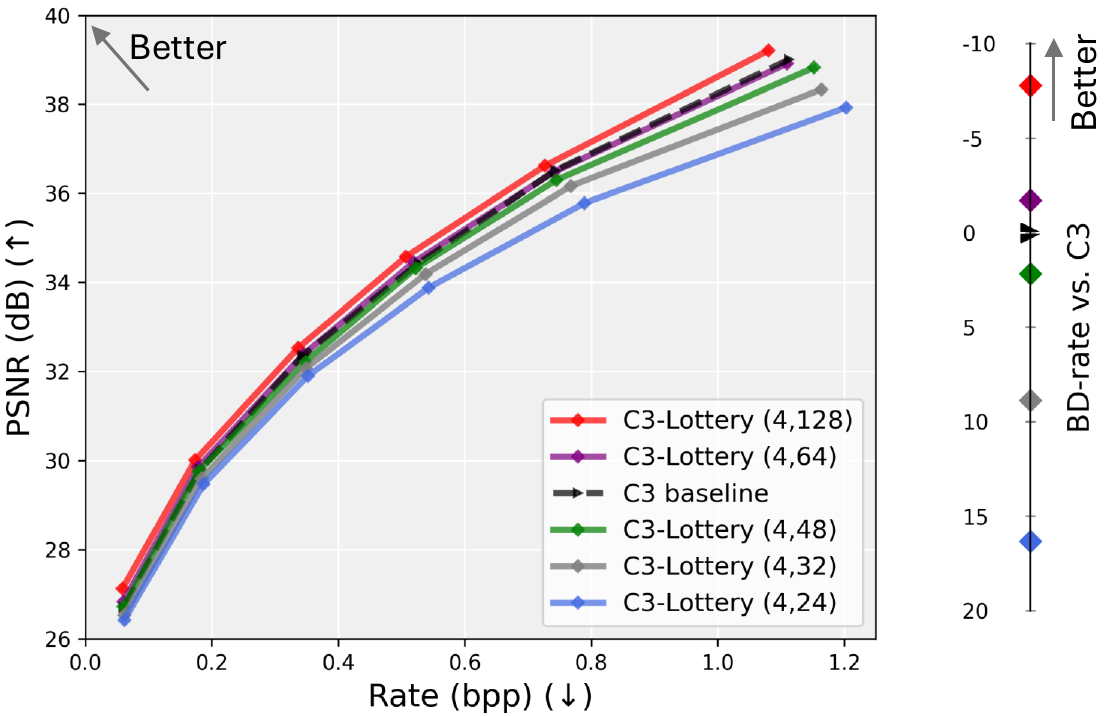}
        \label{op_psnr_rate}
    }%
    \hspace{+1pt}
    \subfloat[]{
        \includegraphics[width=0.22\linewidth]{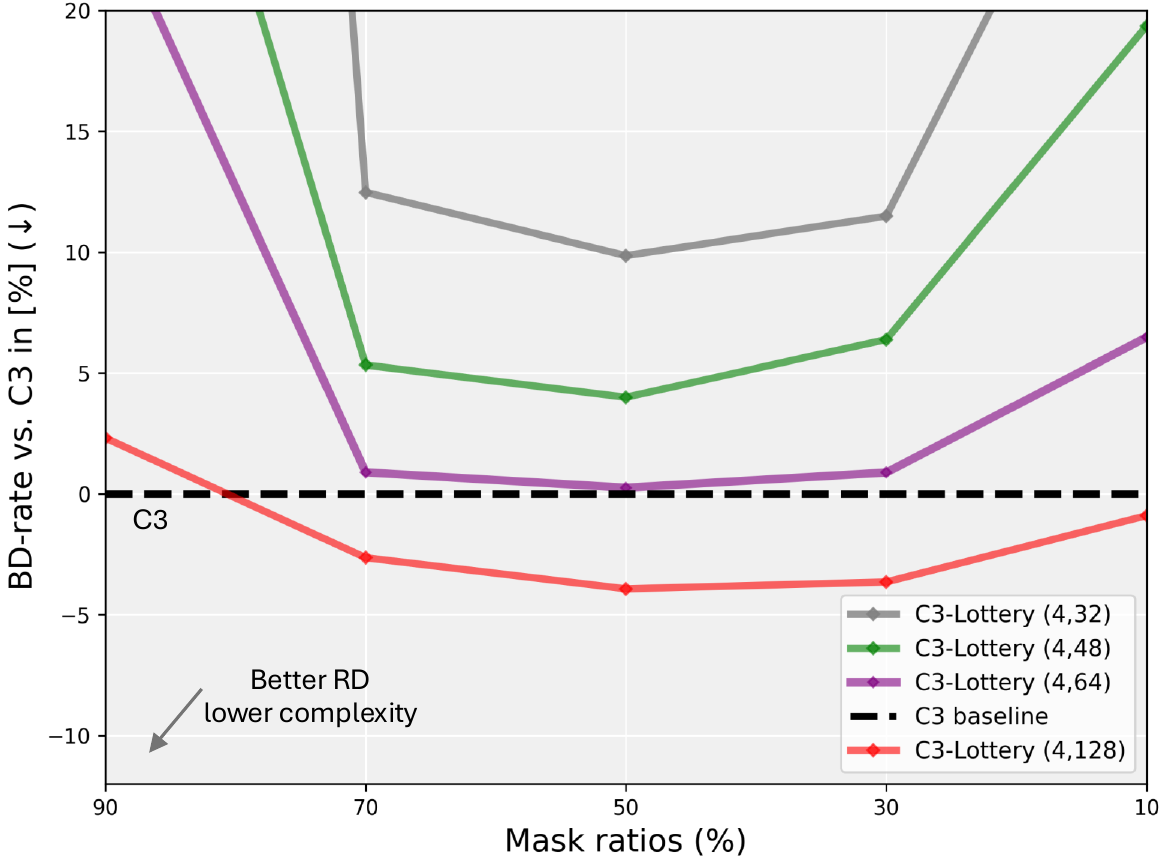}
        \label{op_bd_mask_ratio}
    }%
   \hspace{+1pt}
    \subfloat[]{
        \includegraphics[width=0.22\linewidth]{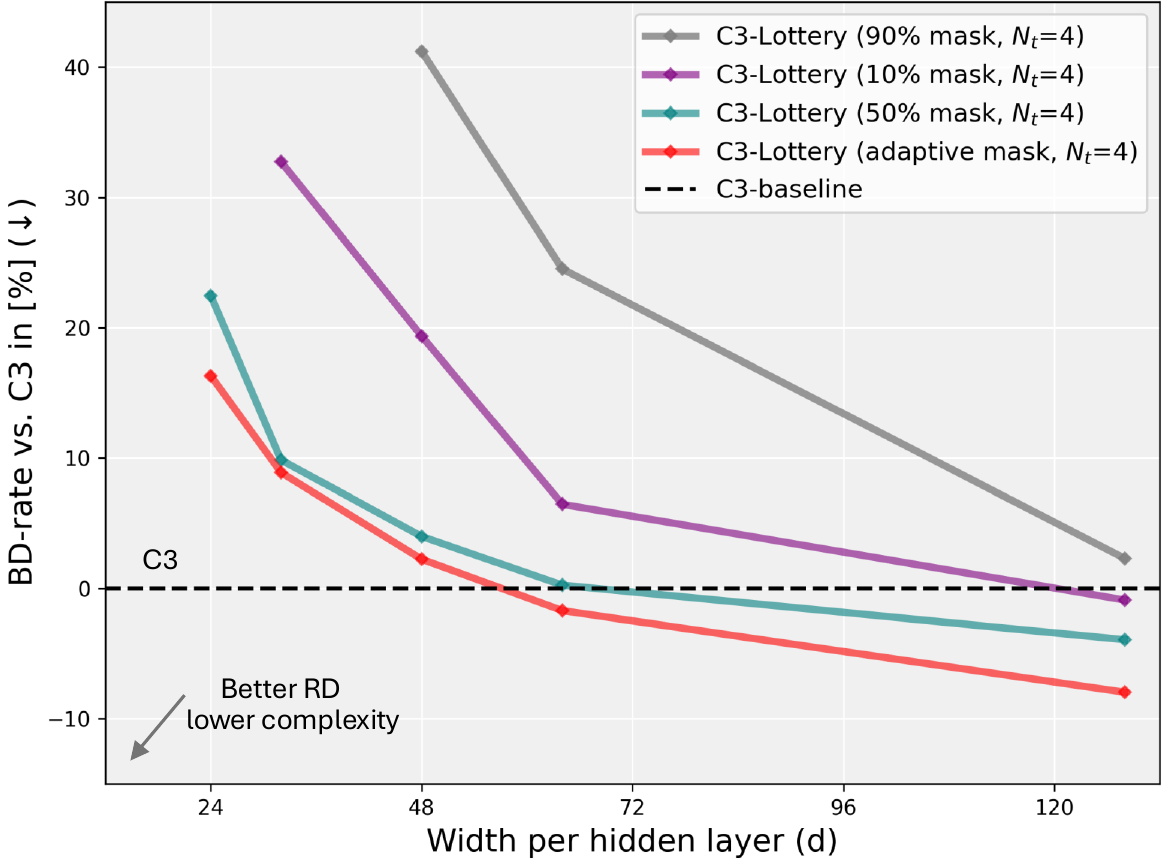}
        \label{op_bd_width}
    }%
    \hspace{+1pt}
    \subfloat[]{
        \includegraphics[width=0.22\linewidth]{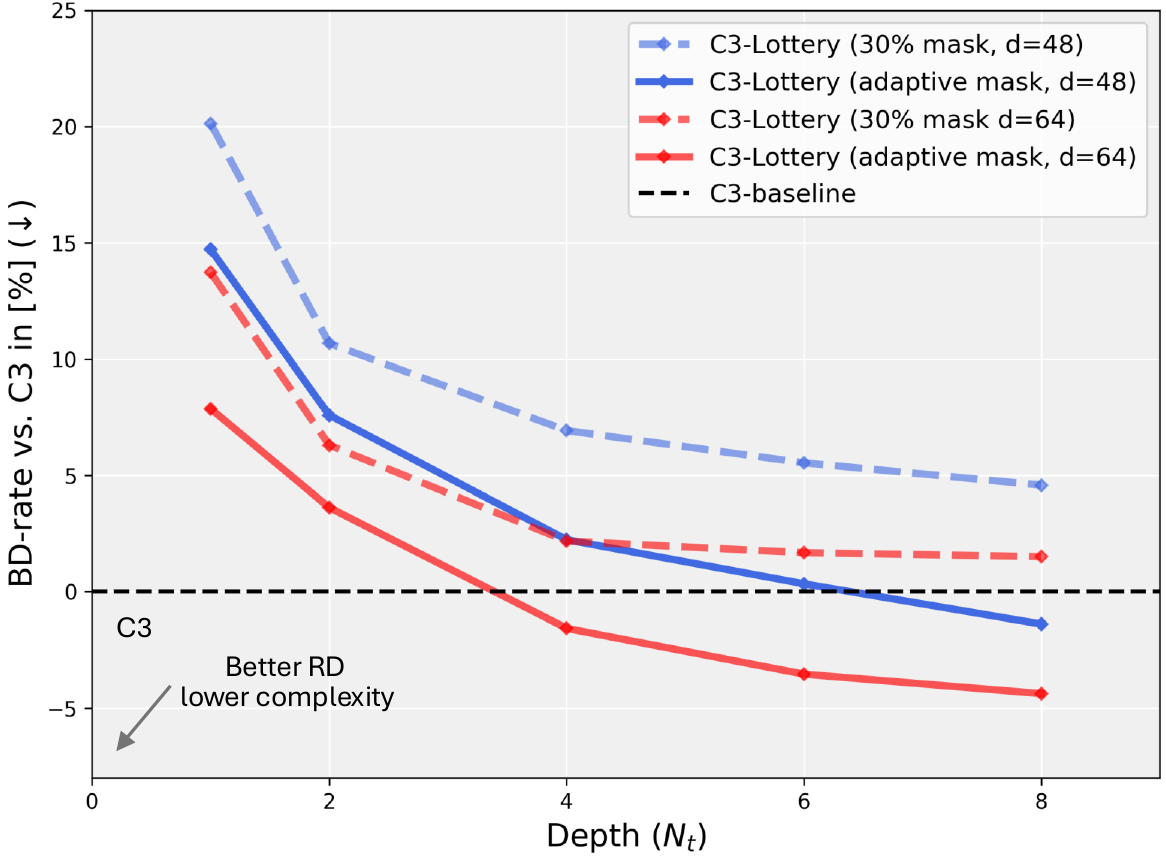}
        \label{op_bd_depth}
    }%
    \caption{Experimental verification of the \textit{lottery codec hypothesis}, where C3-lottery $(N_t,d)$ refers to the scheme using an over-parameterized network with $N_t$ hidden layers and $d$ dimensions per layer. (a) RD curve and BD rate for different over-parameterization configurations. (b) BD-rate versus different mask ratios. (c)-(d) BD-rate over varying width and depth configurations.}
    \label{ab_over_para}
\end{figure*}


\paragraph{Rewind modulation mechanism.}
In practice, directly searching for a well-performing subnetwork from $g_\mathbf{W_0}$ requires a deep, wide random network, leading to high compression and computational costs. To address this, we introduce ModNet $f_{\mathbf{\theta}}$ to generate modulations to simplify subnetwork search and improve the RD performance. 
Specifically, $f_{\mathbf{\theta}}$ takes the quantized latent modulation $\mathbf{\hat{z}}$ as input, where $\mathbf{{z}}\triangleq\{\mathbf{{z}_1},\mathbf{{z}_2},\ldots,\mathbf{{z}_{L}}\}$ comprises $L$ learnable multi-resolution vectors, 
with each $\mathbf{{z}_i}\in \mathbb{Z}^{\frac{HW}{4^{i-1}}}$ being a learnable vector. In $f_{\mathbf{\theta}}$, $\mathbf{\hat{z}}$ is first upsampled using transpose convolutional operations as in \cite{blard2024overfitted}, producing $\mathbf{{U_0}}=[\mathbf{{u}_{1}};\mathbf{{u}_{2}};\ldots;\mathbf{{u}_{L}}]\in\mathbb{Z}^{L\times HW}$, where each $\mathbf{{u}_i}\triangleq\textit{Upsample}(\mathbf{\hat{z}_i})\in \mathbb{Z}^{HW}$. The resultant $\mathbf{{U_0}}$ is then processed through convolutional layers as: $\mathbf{U_i}=f_{\mathbf{\theta}}^{(i)}(\mathbf{U_{i-1}})$, $i=1,\ldots,L_t-1$. Here, $f_{\mathbf{\theta}}^{(i)}$ and $\mathbf{U_i}$ denote the operation and output of the $i$-th layer of ModNet, respectively. 

Inspired by \cite{mehta2021modulated,perez2018film,zhou2019deconstructing,anonymous2024find} and to leverage the network structure to encode the source image, we propose a rewind strategy to modulate the synthesis process by concatenating compensational structural information in a rewind fashion. 
The modulation vector for the $i$-th layer of the synthesis network $g_{\tau\odot \mathbf{W_0}}$ is defined as:
\begin{equation}
    \mathbf{M_i}=\textit{Concatenate}(\mathbf{U_{L_t-1}},\mathbf{U_{L_t-2}},\ldots,\mathbf{U_{L_t-i}}),
\end{equation}
where vectors $\mathbf{U_{i}}$'s from ModNet are concatenated in reverse order to facilitate the search for a high-performing subnetwork.

The modulation operation concatenates $\mathbf{M_i}$ with the intermediate layer output of $g_{\tau\odot \mathbf{W_0}}$. Consequently, the output of each masked linear layer after modulations is given by:
\begin{equation}
\begin{aligned}
    \mathbf{G_i} &= m(\mathbf{F_i})= \textit{Concatenate}(\mathbf{M_i}, \mathbf{F_i}),
\end{aligned}
\end{equation}
where $\mathbf{F_i}$ represents the output of the $i$-th layer of $g_{\tau\odot \mathbf{W_0}}$, and $\mathbf{G_{i}}$ is the output after corresponding modulation operations. Intuitively, this concatenation enriches the structure of the SuperMask network with both sign and magnitude information, allowing for the reactivation of features in deeper layers while preserving high-level features. Note that the proposed LotteryCodec scheme with its rewind modulation mechanism serves as a general framework, allowing for alternative modulation operations, such as FiLM \cite{perez2018film}, which are discussed in Appendix \ref{ap_mod}.
\begin{figure*}[t]
    \centering
    \subfloat[]{
        \includegraphics[width=0.322\linewidth]{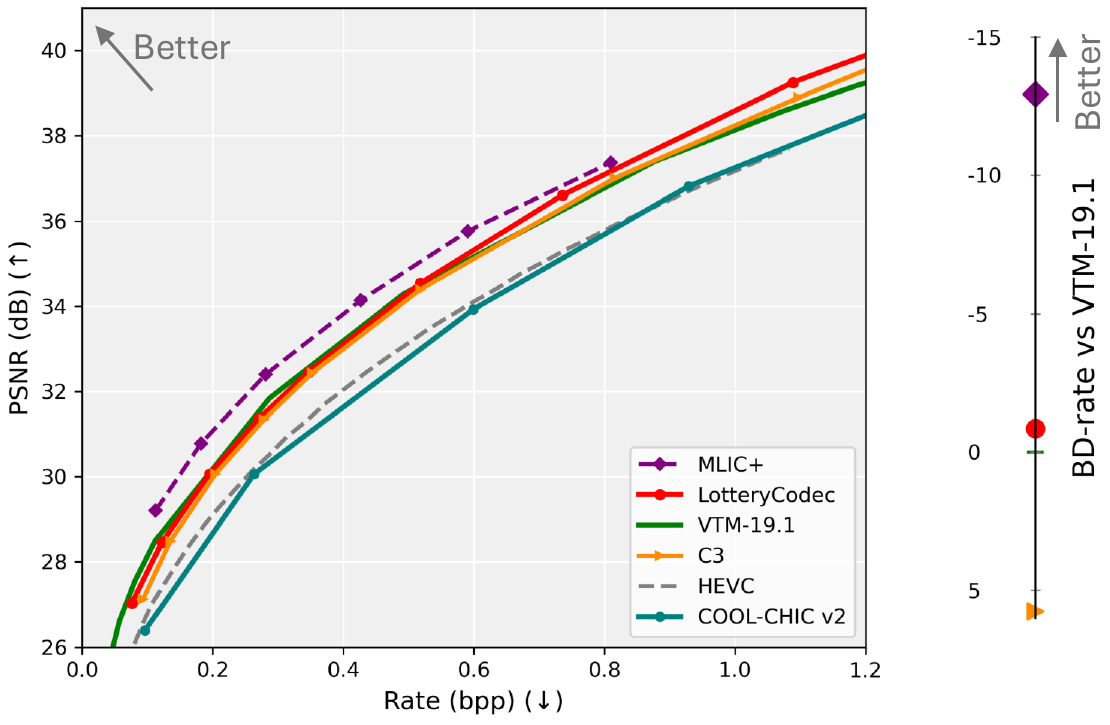}
        \label{psnr_rate_kodak}
    }%
\hspace{+1pt}
     \subfloat[]{
        \includegraphics[width=0.324\linewidth]{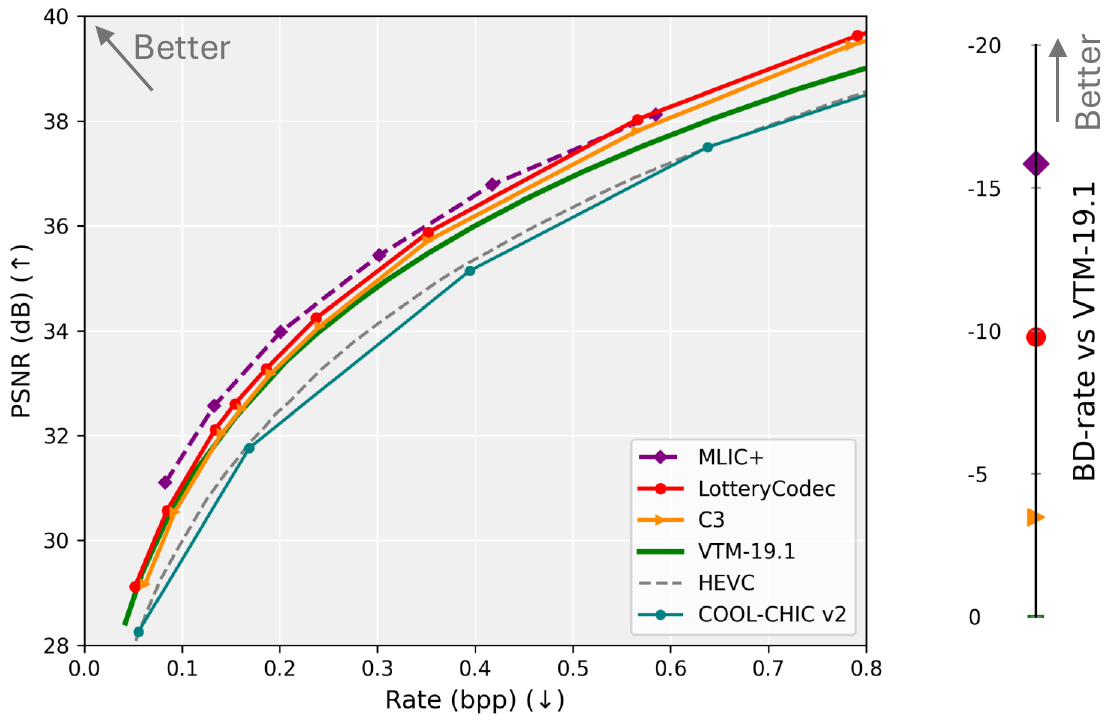}
        \label{psnr_rate_clic}}
       \hspace{+1pt}
   \subfloat[]{
        \includegraphics[width=0.282\linewidth]{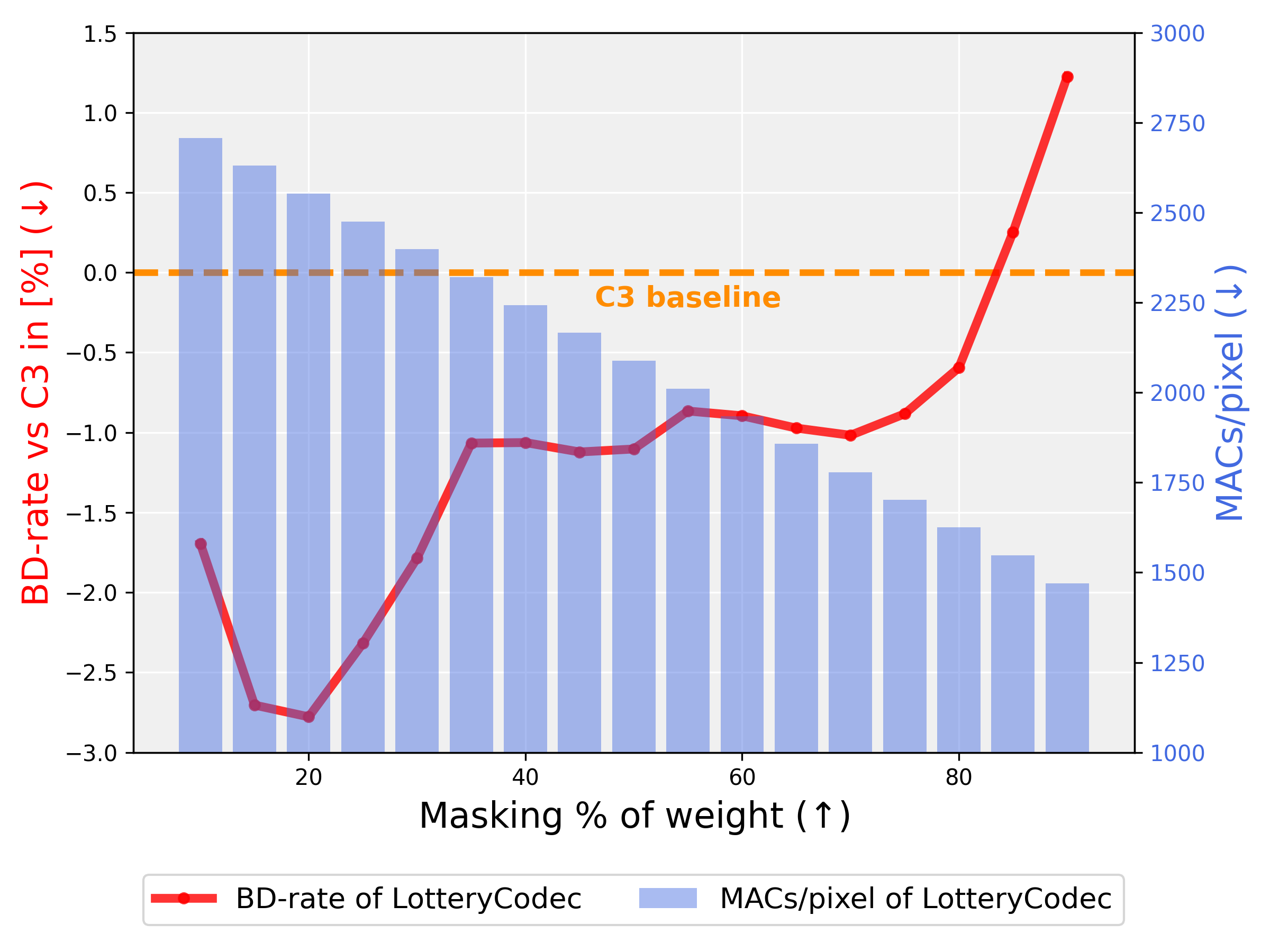}
        \label{op_bd_spar_macs}
    }%
   \caption{Performance of LotteryCodec and other schemes. (a) RD curve and BD rate on Kodak dataset. (b) RD curve and BD rate on CLIC2020 dataset. (c) BD-rate and decoding complexity across different mask ratios on Kodak dataset.}
    \label{main_result}
\end{figure*}


\section{Experimental results}
We evaluate our model on the Kodak ($24$ images) \cite{kodak_dataset} and CLIC2020 ($41$ images) \cite{clic_2020} datasets. The mask ratio is selected from $[0.1, 0.9]$. 
LotteryCodec is compared with classical codecs (VTM-19.1 \cite{bross2021developments}, HEVC \cite{sullivan2012overview}), AE-based neural codecs (CST \cite{cheng2020learned}, EVC \cite{wang2023evc}, MLIC+ \cite{jiang2023mlic}), and other overfitted image codecs (including C3 \cite{kim2024c3} and COOL-CHIC \cite{leguay2023low}). The peak signal-to-noise ratio (PSNR) on RGB channels and BD-rate \cite{gisle2001calculation} are used to evaluate the RD performance. The evaluation details and additional experiments, including ablation study and visualizations, are presented in the Appendix \ref{ap_vis}.

\subsection{Verification of the \textit{lottery codec hypothesis}}
\label{exp_verify_hypothesis}
We first validate the \textit{lottery codec hypothesis} through a series of experiments on the first $10$ images of the Kodak dataset. 
We implement the C3 scheme with a $16$-dimensional ARM as the baseline. For comparison, we search for a subnetwork within a randomly initialized network and replace the well-trained synthesis network in C3 with this subnetwork, keeping all other components unchanged. This method is referred to as the C3-Lottery scheme.
See {Appendix \ref{over_para_details}} for more details.

Fig.~\ref{ab_over_para} presents the RD trade-off across varying network depths and widths. The results demonstrate that when the randomly initialized network is sufficiently over-parameterized, the C3-Lottery scheme successfully identifies a subnetwork and an associated latent representation that achieves the same level of distortion as C3, but at a lower bitrate. 
In particular, Fig. \ref{op_psnr_rate} presents PSNR versus the rate contributed by $\hat{\mathbf{z}}$, demonstrating that RD performance (in terms of the PSNR-rate curve and BD-rate) improves as the network {width} increases. In particular, when depth $N_t=4$, the C3-Lottery scheme matches or even surpasses the performance of well-trained C3 baselines for width $d\ge 64$. Additional results for different depth settings can be found in Appendix \ref{over_para_details}. Fig. \ref{op_bd_mask_ratio} illustrates the impact of the mask ratio on performance, revealing that the optimal LotteryCodec is achieved at a mask ratio of approximately $50\%$. Intuitively, this occurs because a $50\%$ mask ratio maximizes the entropy of the network structure, allowing more information to be encoded into the structure. Figs. \ref{op_bd_width} and \ref{op_bd_depth} indicate that increasing both the depth and the width of the random network can significantly enhance RD performance. Furthermore, employing adaptive mask ratios for different images could further boost performance.

\subsection{RD performance of the LotteryCodec}
We then compare the RD performance of LotteryCodec with the existing methods. As shown in Fig.~\ref{psnr_rate_kodak}, LotteryCodec outperforms VTM and significantly surpasses other overfitted codecs on the Kodak dataset, achieving BD-rate reductions of $-6.4\%$ over C3 and $-3.73\%$ over its adaptive variant, C3-adapt (relative to VTM-19.1). 
One of the substantial performance gains occurs in the low-bpp regime, where network parameters dominate the compression rate in an overfitted codec. By contrast, LotteryCodec achieves a lower rate by compressing a binary mask instead of real-valued parameters, resulting in a superior RD performance (more details are provided in Appendix \ref{ap_bd_complexity}). Additional MS-SSIM evaluations are provided in the appendix (see Fig.~\ref{fig:MS_SSIM}), where LotteryCodec closely approaches ELIC and has a $+10.72\%$ BD-rate gap (versus VTM-19.1) compared to MLIC+.

Fig. \ref{psnr_rate_clic} shows the results on the CLIC2020 dataset, where LotteryCodec demonstrates an even greater advantage. In particular, the performance gain over VTM-19.1 reaches up to $-9.79\%$ (see Appendix \ref{ap_bd_complexity} for details). Consistent improvements are also observed from Fig. \ref{psnr_rate_clic} over C3 ($-5.69\%$ BD-rate), C3-adpt ($-2.90\%$ BD-rate) and COOL-CHIC v2 ($-25.57\%$ BD-rate), respectively. From experiments on both the Kodak and CLIC2020 datasets, the LotteryCodec scheme demonstrates state-of-the-art RD performance among current overfitted codecs. \textit{To the best of our knowledge, LotteryCodec is the first neural codec to surpass VTM in RD performance while maintaining low decoding complexity, establishing it as the state-of-the-art for single-image compression.} Like other overfitted codecs, LotteryCodec still falls short of state-of-the-art AE-based neural codecs such as MLIC+ in RD performance. But it achieves a BD-rate over VTM that is comparable to MLIC, with only a $+2.33\%$ gap. As shown in Fig. \ref{fig:complexity_clic}, LotteryCodec demonstrates a significant advantage over MLIC+ in terms of decoding complexity, requiring two orders of magnitude fewer decoding operations. Further discussion on complexity is provided later.

\subsection{The mask ratio and decoding complexity}
The impact of the mask ratio is examined in Fig. \ref{op_bd_spar_macs}, which illustrates the performance of LotteryCodec under different mask ratios. As expected, the decoding complexity decreases linearly as the mask ratio increases, since masking more weights reduces the overall computation required by the network. However, RD performance does not follow a monotonic trend with respect to the mask ratio. Interestingly, with the introduction of the rewind modulation, the optimal mask ratio for LotteryCodec to achieve the best BD-rate is around $20\%$, different from $50\%$ observed in Fig. \ref{ab_over_para}. 
Intuitively, decreasing the mask ratio from $50\%$ to $20\%$ significantly reduces the number of possible subnetworks, demonstrating that rewind modulation effectively simplifies the subnetwork search process while maintaining strong compression performance.

Additionally, Fig. \ref{op_bd_spar_macs} demonstrates that LotteryCodec can flexibly balance RD performance and decoding complexity, allowing for adaptable trade-offs based on specific requirements. Notably, even with a high mask ratio of $80\%$, which results in low decoding complexity, LotteryCodec still outperforms the C3 baseline scheme. More detailed ablation studies and additional experiments on varying mask ratios are presented in Appendix \ref{ap_bd_complexity}

A more comprehensive comparison of decoding complexity between LotteryCodec and other neural codecs is presented in Fig. \ref{fig:complexity_clic} and Fig. \ref{bd_copmlexity_kodak}. In particular, compared to most AE-based schemes, LotteryCodec achieves similar or superior RD performance with at least an order of magnitude fewer MACs. For more advanced codecs like ELIC and MLIC+, while LotteryCodec does not exceed their RD performance, it reduces MACs by over two orders of magnitude while maintaining acceptable quality. Compared to other overfitted image codecs, such as the COOL-CHIC/C3 family, LotteryCodec achieves better RD performance with lower decoding complexity. 

Notably, the current figure shows the theoretical minimum decoding complexity, excluding masked operations. This lower bound can be approached with sparsity-aware libraries (e.g., TVM \cite{chen2018tvm}, cuSparse \cite{naumov2010cusparse}, DeepSparse \cite{pmlr-v119-kurtz20a}) on compatible hardware. 
For a comprehensive analysis, we also report real coding times using structured pruning (see Tables~\ref{tab:encoding_time} and~\ref{tab:encoding_time_across_resolution} in the appendix). Additionally, we provide both theoretical upper and lower bounds on decoding complexity, corresponding to unpruned and active operations, respectively, with practical complexity lying in between. As shown in Fig.~\ref{fig:ab_flexible_BD}, even without pruning, LotteryCodec outperforms C3 in BD-rate with similar complexity.

\section{Conclusion and future work}
This paper introduces and validates the \textit{lottery codec hypothesis}, which proposes a new paradigm for image compression: encoding images into structures of randomly initialized networks. Building on this hypothesis, we propose LotteryCodec, a novel overfitted codec that compresses an image into modulation vectors and a binary mask for an over-parameterized, randomly initialized network. The proposed LotteryCodec achieves state-of-the-art RD performance among existing overfitted codecs while maintaining adaptive and low decoding complexity. Our work advances the field of overfitted image compression by addressing the critical challenge of achieving high compression efficiency with minimal computational cost.  
\paragraph{\textcolor{black}{Limitations.}} LotteryCodec’s low and flexible decoding cost is particularly beneficial in multi-user streaming scenarios, where encoding can be done once and offline, to support many users in decoding the same content. While high encoding complexity remains a key bottleneck for all overfitted codecs, including ours, several potential acceleration strategies exist, such as meta-learning, mixed-precision training, and neural architecture search. In particular, LotteryCodec also opens the door to parallel encoding of overfitted codecs by reparameterizing distinct network learning processes into a batch of mask learning processes.

\paragraph{\textcolor{black}{Future work.}} As a novel paradigm in image compression, LotteryCodec opens several avenues for future research. First, its performance and efficiency could be further optimized by incorporating advanced strategies, such as those proposed in \cite{kim2024c3,ladune2024cool}. 
Second, LotteryCodec can be extended as a flexible alternative for video coding. By sharing modulation across adjacent {groups of frames} (GoF) and applying distinct masks, it can also encode temporal information into the network structure. Moreover, video coding enables adaptive mask ratio selection across GoF, offering greater flexibility in both computational complexity and rate control. Its low decoding complexity positions LotteryCodec as a promising approach for real-world neural compression applications, enabling efficient deployment in resource-constrained environments.

\section*{Impact Statement}
This paper presents work whose goal is to advance the field of Machine Learning. There are many potential societal consequences of our work, none of which we feel must be specifically highlighted here.

\section*{Acknowledgments}
This work received funding from the UKRI for the projects INFORMED-AI (EP/Y028732/1) and AI-R (ERC Consolidator Grant, EP/X030806/1).
\bibliography{example_paper}
\bibliographystyle{icml2025}

\newpage
\appendix
\onecolumn
\section*{\Large{Appendix}}
\section{Quantization and entropy coding methods}
\label{quantization_entropy_coding}
For compression, the latent modulation $\mathbf{z}$ and network parameters $\mathbf{\theta}$, $\mathbf{\psi}$ are quantized into $\mathbf{\hat{z}}$, $\mathbf{\hat{\theta}}$ and $\mathbf{\hat{\psi}}$, respectively, and subsequently entropy-coded into a bitstream, using standard methods, as in \cite{kim2024c3,ladune2024cool}. More details can be seen in the Table \ref{tab:model_parameter}.
\subsection{Latent modulation}
\paragraph{Quantization.}
Similar to \cite{kim2024c3}, we adopt a two-stage quantization-aware optimization approach for optimizing $\mathbf{z}$. During the training stage, $\mathbf{z}$ is learned in a continuous space for discrete optimization, with quantization approximated using Kumaraswamy noise. This soft-rounding technique ensures that the quantization process remains differentiable with respect to $\mathbf{z}$. In the inference stage, uniform quantization with hard-rounding is applied to $\mathbf{z}$ as:
\begin{equation}  
\mathbf{\hat{z}}=
             \begin{cases}
             \mathcal{S}_{T}(\mathbf{z})+\mathbf{u}_{kum}, &\text{Training Stage I}\\  
             {Q}(\mathbf{z}), & \text{Training Stage I \& Inference Stage},\\  
             \end{cases}   
\end{equation} 
where $\mathcal{S}_{T} $ denotes soft-rounding operation with temperature $T$, $Q$ represents hard-rounding, and $\mathbf{u}_{kum}$ is the Kumaraswamy noise term. The temperature and noise strengths are controlled to shape the Kumaraswamy distribution, transitioning from a peaked form (low noise) at the beginning of the training stage to a uniform distribution by its end.

\paragraph{Entropy coding.} Similar to \cite{balle2018variational,ladune2023cool,NEURIPS2018_53edebc5}, we introduce a factorized auto-regressive model $r_{\mathbf{\psi}}$ to estimate the distribution of $\mathbf{\hat{z}}$, which is necessary in the entropy coding algorithm. The distribution of each latent element of ${\hat{z}_{i,j}}$ (the $j$-th element of $\mathbf{z}_i$) is conditioned on ${C}$ spatially neighboring elements $\mathbf{c_{i,j}}\in \mathbb{Z}^{C}$ as:
\begin{equation}
    p_{\mathbf{\psi}}(\hat{\mathbf{z}})=\prod_{i,j}p_{\mathbf{\psi}}(\hat{z}_{i,j}|\mathbf{c_{i,j}}),
\end{equation}
where $p_{\mathbf{\psi}}(\hat{\mathbf{z}})$ is modeled by an integrated Laplace distribution as
\begin{equation}
    p_{\mathbf{\psi}}(\hat{z}_{i,j}|\mathbf{c_{i,j}})=\int_{\hat{z}_{i,j}-0.5}^{\hat{z}_{i,j}+0.5}g(z)dz.
    \label{eq_laplace}
\end{equation}
The expectation and scale parameters are estimated via context elements as $g\sim \mathcal{L}(\mu_{i,j},\sigma_{i,j})$, where $\mu_{i,j},\sigma_{i,j}=r_{\mathbf{\psi}}(\mathbf{c_{i,j}})$.


With this estimated distribution of $\hat{\mathbf{z}}$, the range coding algorithm is used (range-coder in PyPI), as \cite{kim2024c3} \footnote{An optimized C++ implementation is open-sourced on COOL-CHIC project page.}, to compress $\hat{\mathbf{z}}$. The resulting rate contributed by $\mathbf{\hat{z}}$ is then given by:
\begin{equation}
\begin{aligned}
    R(\mathbf{\hat{z}})
    &=-\log_2 p_{\mathbf{\psi}}(\mathbf{\hat{z}})= - \sum_{i,j}\log_2 p_{\mathbf{\psi}}(\hat{z}_{i,j}|\mathbf{c_{i,j}}).
\end{aligned}
\label{eq_rate_y}
\end{equation}

\subsection{Model compression}
The parameters of ModNet and ARM are essential for decoding and are therefore compressed. Specifically, we first quantize $\mathbf{\psi},\mathbf{{\theta}}$ using a scalar quantizer ${Q}(\cdot,\Delta)$ with step size $\Delta$ as:
\begin{equation}
    \mathbf{\hat{\theta}}={Q}(\mathbf{\theta},\Delta_\mathbf{\theta}), \text{and} \quad\mathbf{\hat{\psi}}={Q}(\mathbf{\psi},\Delta_\mathbf{\psi}).
\end{equation}
The quantized parameters $\hat{{\theta}}$ and ${\hat{\psi}}$ are then entropy-coded, where the discrete distribution of each quantized parameter is modeled by a continuous Laplace distribution. 
Specifically, the probability of a quantized parameter from $\mathbf{\theta}$ (similarly for $\mathbf{\psi}$) is given by:
\begin{equation}
    p(\hat{{\theta}}_i)=\int_{\hat{{\theta}}_i-0.5}^{\hat{{\theta}}_i+0.5}g(\theta)d\theta,  \quad \text{with} \quad g\sim \mathcal{L}(0,\text{stddev}(\mathbf{\hat{\theta}})),
\end{equation}
where $g$ follows a Laplace distribution with zero mean and a standard deviation stddev ($\hat{\theta}$) of the quantized parameters.

Then, the total rate contribution from both ARM and ModNet can be expressed as:
\begin{equation}
\begin{aligned}
    R_{\text{MLP}}
    =R_{\mathbf{\hat{\theta}}}+R_{\mathbf{\hat{\psi}}}=\sum_{i}-\log_2 p(\hat{\theta}_{i})+\sum_{i}-\log_2 p(\hat{\psi}_{i}).\\
\end{aligned}
\label{eq_rate_mlp}
\end{equation}
A greedy search over quantization steps selects optimal $\Delta_{\theta}$ and $\Delta_{\psi}$ by minimizing the following rate-distortion cost:
\begin{equation}
    \min_{\Delta_{\psi},\Delta_{\theta}} D(\mathbf{S},g_{\mathbf{\tau}\odot \mathbf{W_0}}(f_{\mathbf{\hat{\theta}}}(\mathbf{\hat{z}}),\mathbf{x}))+\lambda R,
\end{equation}
where $R$ is the same as that in Eqn. \eqref{eq: RD-LTH}.

\section{Implementation details.}
\subsection{Baseline choices}
LotteryCodec is compared against classical codecs, including VTM \cite{bross2021developments} and HEVC \cite{sullivan2012overview}, as well as autoencoder-based neural codecs: CST \cite{cheng2020learned} (a competitive neural codec), EVC \cite{wang2023evc} (optimized RD performance with low decoding complexity), and MLIC+ \cite{jiang2023mlic} (one of the state-of-the-art neural codec). Additionally, we compare with overfitted INR-based codecs such as C3 \cite{kim2024c3} (state-of-the-art overfitted image codec) and COOL-CHICv2 \cite{leguay2023low} (an optimized version of COOL-CHIC version). We measure PSNR on RGB channels and quantify RD performance using the BD-rate metric \cite{gisle2001calculation}. The baseline results were obtained using their official implementations or directly using the reported results (C3, MLIC+) from their papers. For VTM, we use the CompressAI library~\cite{begaint2020compressai} for an updated VTM-19.1 (YUV $10$ bits) version, where code and datapoints are provided on our project website.


\subsection{Datasets}
Unless specified otherwise, we use the Kodak dataset and CLIC2020 professional validation sets for evaluation. Specifically, the Kodak dataset consists of $24$ images, each with a resolution of $512\times 768$. The CLIC2020 professional validation set includes $41$ images with resolutions ranging from $439\times 720$ to $1370\times 2048$.

\subsection{Model architecture}
We provide the detailed architecture setting for the proposed LotteryCodec scheme in Table \ref{tab:model_parameter}. 

For the ARM model with $c$ contextual elements as input, denoted as ARM-$c$ model, there are three linear or $1\times1$ convolutional layer, followed by GELU activation functions, with input and output dimension given as $(c\times c)\rightarrow \text{GELU} \rightarrow (c\times c)\rightarrow \text{GELU}\rightarrow  (c\times 2)$. For the proposed LotteryCodec, we employ $c\in\{8,16,24,32\}$.

The ModNet model, using $L$-dimensional latent modulation as the input, comprising $L_t-1$ layers ($L_t=4$ and $L=7$ in this paper), the input and output dimension are given as: $(7\times d)\rightarrow \text{GELU} \rightarrow (d\times 3)\rightarrow \text{GELU}\rightarrow  (3\times 3)$. For the proposed LotteryCodec, we employ $c\in\{32,48\}$ for Kodak and CLIC2020 dataset. 

For the SuperMask model using pixel coordinates as the input, comprising $L_t=4$ layers, the input and output dimension are given as: $(2\times 32)\rightarrow \text{GELU} \rightarrow ([32+3] \times 24)\rightarrow \text{GELU}\rightarrow  ([24+3+3] \times 16) \rightarrow  ([16+3+3+d] \times 3)\rightarrow \text{Tanh}$. 
\label{improved_model}

{To optimize the RD performance, we can further add a $3\times3$ convolutional operation in ModNet: $(7\times d)\rightarrow \text{GELU} \rightarrow (d\times 3)\rightarrow \text{GELU}\rightarrow  (3\times 3)\rightarrow \text{GELU}\rightarrow  (3\times 3)$, where the last two output are concatenated as the first modulation input. As a result, the SuperMask mapping becomes: $(2\times 32)\rightarrow \text{GELU} \rightarrow ([32+3+3] \times 24)\rightarrow \text{GELU}\rightarrow  ([24+3+3+3] \times 16) \rightarrow  ([16+3+3+3+d] \times 3)$. This additional modification can be omitted if lower complexity is prioritized.}


\subsection{over-parameterization experiments}
\label{over_para_details}
This section outlines the detailed experimental settings of Section \ref{exp_verify_hypothesis}. 

\paragraph{Datasets.} Section. \ref{exp_verify_hypothesis} trains over-parameterized models with varying mask ratios $\{10\%,20\%,30\%,50\%,70\%,90\%\}$ and different $\lambda\in \{2e-2,1e-2,5e-3,1e-3,5e-4,2e-4\}$ to compute the BD rate, which yields $36$ models for each image. Considering extensive experiments, we use the first $10$ images of Kodak dataset to compare the average performance, which indicates that each point in the figure corresponds to an average performance of $360$ well-trained INR models.

\begin{figure*}[t]
    \centering
\includegraphics[width=0.8\linewidth]{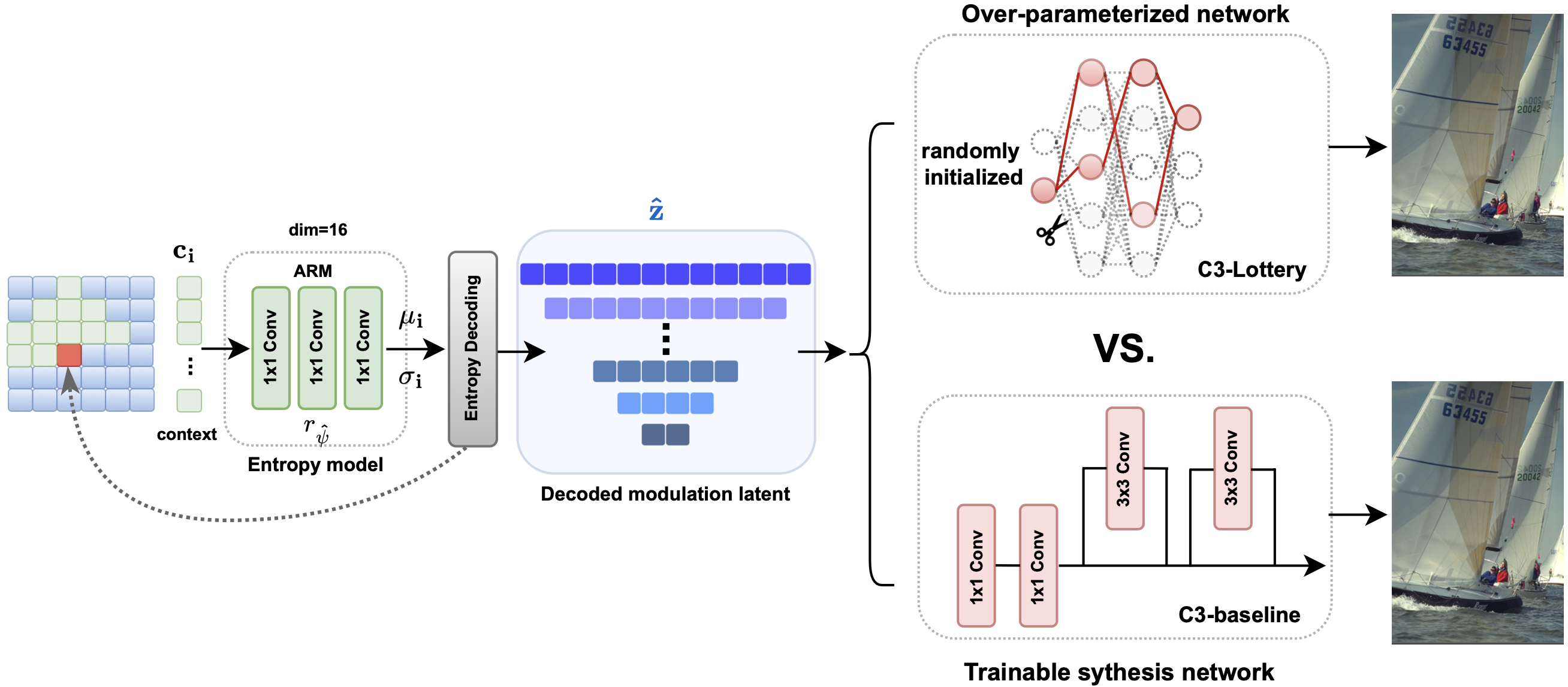}
    \caption{Illustration of the experiments in Section \ref{exp_verify_hypothesis}: the synthesis network is replaced with a randomly initialized over-parameterized network, where only a binary mask is learned, while all other components remain unchanged for a fair comparison.}
\label{fig:over_exp_fig_illustration}
\end{figure*}

\begin{table*}[t]
\centering
\begin{tabular}{lcc}
\toprule
\textbf{Hyper parameter} & \textbf{Initial values} & 
\textbf{Final values} \\
\hline
\textbf{Values of $\lambda$} & \multicolumn{2}{c}{$ \{2e^{-2},1e^{-2},5e^{-3},1e^{-3},5e^{-4},2e^{-4}\}$}\\
\hline
\multicolumn{3}{c}{\textbf{Quantization – Stage I}} \\ 
\hline
Number of encoding steps & \multicolumn{2}{c}{$10^5$}  \\
Learning rate $\beta$ & {$10^{-2}$} & $0$ \\
Scheduler for learning rate & \multicolumn{2}{c}{Cosine scheduler}\\
Temperature $T$ for soft rounding & $0.3$ &  $0.1$ \\
Noise strength $\alpha$ for Kumaraswamy noise & $2.0$ & $1.0$ \\
Scheduler for Soft-rounding and Kumaraswamy noise & \multicolumn{2}{c}{Linear scheduler}\\
\hline
\multicolumn{3}{c}{\textbf{Quantization – Stage II}} \\ 
\hline
Number encoding steps & \multicolumn{2}{c}{$10^4$}  \\
Learning rate & $10^{-4}$ & $10^{-8}$ \\
Decay learning rate if loss has not improved for this many steps &  \multicolumn{2}{c}{$40$}\\
Decay factor & \multicolumn{2}{c}{$0.8$} \\
Temperature $T$ for soft rounding & \multicolumn{2}{c}{$10^{-4}$} \\
\hline
\multicolumn{1}{l}{\textbf{Architecture – Latent modulations}} & \multicolumn{2}{c}{\textbf{Values}} \\ 
\hline
Number of latent vectors $L$ & \multicolumn{2}{c}{$7$} \\
Initialization of $\mathbf{z}$ & \multicolumn{2}{c}{$0$} \\
\hline
\multicolumn{1}{l}{\textbf{Architecture – ModNet}} & \multicolumn{2}{c}{\textbf{Values}} \\
\hline
Output channels of the $1 \times 1$ convolutions & \multicolumn{2}{c}{$\{48,3\}$ vs. $\{32,3\}$} \\
Number of $3 \times 3$ residual convolutions & \multicolumn{2}{c}{$\{1\}$ vs $\{2\}$} \\
\hline
\multicolumn{3}{l}{\textbf{Architecture – Entropy model}} \\ 
\hline
Alternative widths of the $3$ layers residual $1 \times 1$ convolutions (ARM-$c$) &   \multicolumn{2}{c}{$c\in\{8 / 16 / 24 / 32\}$} \\
Activation function &   \multicolumn{2}{c}{GELU} \\
Log-scale of Laplace is shifted before $\exp$ & \multicolumn{2}{c}{$4$} \\
Scale parameter of Laplace is clipped to &  \multicolumn{2}{c}{$[10^{-2}, 150]$} \\
\hline
\multicolumn{1}{l}{\textbf{Architecture – SuperMask network}} & \multicolumn{2}{c}{\textbf{Values}} \\
\hline
Output dimensions of MLP layers & \multicolumn{2}{c}{ $\{32,24,16,3\}$} \\
Mask ratios & \multicolumn{2}{c}{ $\{0.1,0.9\}$ with $0.05$ intervals} \\
Initialization of MLP & \multicolumn{2}{c}{FFN initialization} \\       
FFN initialization phase number $P$ & \multicolumn{2}{c}{$32$} \\
FFN initialization low/high frequency number $F$& \multicolumn{2}{c}{$64/64$} \\

Initialization of score matrix $\mathbf{P}$ & \multicolumn{2}{c}{Kaiming uniform initialization} \\
Learning rate $\alpha$ for $\mathbf{P}$ & \multicolumn{2}{c}{$0.1$} \\
Scheduler & \multicolumn{2}{c}{Cosine scheduler}\\
\bottomrule
\end{tabular}
\caption{Hyper-parameter settings}
\label{tab:model_parameter}
\end{table*}
\paragraph{Model architecture} 
As illustrated in Fig. \ref{fig:over_exp_fig_illustration}, we replace the synthesis network of C3 with an over-parameterized network, denoted as C3-Lottery. We present architecture with different over-parameterization levels, as in Table \ref{table:conv_models_go_wider} and Table \ref{table:conv_models_go_deeper}.

{For a fair comparison, we employ the same ARM model for the C3} in Section \ref{exp_verify_hypothesis} as the target network. To ensure a fair evaluation of our hypothesis, only the synthesis network is replaced while all other components are kept unchanged for both schemes, as seen in Fig. \ref{fig:over_exp_fig_illustration}.

\paragraph{\textcolor{black}{Fast implementation.} }
Given that LotteryCodec requires adaptive adjustments to mask ratios and model architectures across images, which may incur extra computational costs, we provide a recommended configuration in Table \ref{tab:fast_model_parameter} that achieves near state-of-the-art performance with less adaptation and faster training.

\begin{table*}[t]
\centering
\begin{tabular}{lcc}
\toprule
\multicolumn{1}{l}{\textbf{Architecture – ModNet}} & \multicolumn{2}{c}{\textbf{Values}} \\
\hline
Output channels of the $1 \times 1$ convolutions & \multicolumn{2}{c}{$\{48,3\}$} \\
Number of $3 \times 3$ residual convolutions & \multicolumn{2}{c}{$\{2\}$} \\
\hline
\multicolumn{3}{l}{\textbf{Architecture – Entropy model}} \\ 
\hline
Alternative widths of the $3$ layers residual $1 \times 1$ convolutions (ARM-$c$) &   \multicolumn{2}{c}{$c\in\{16 / 24\}$} \\
\hline
\multicolumn{1}{l}{\textbf{SuperMask network}} & \multicolumn{2}{c}{\textbf{Values}} \\
\hline
Mask ratios & \multicolumn{2}{c}{ $\{0.15,0.4\}$ with $0.05$ intervals} \\
\bottomrule
\end{tabular}
\caption{Fast implementation of LotteryCodec} 
\label{tab:fast_model_parameter}
\end{table*}

\begin{figure*}[!h]
    \centering
       \hspace{+0.5pt}
    \subfloat[]{
        \centering
    \includegraphics[width=0.47\linewidth]{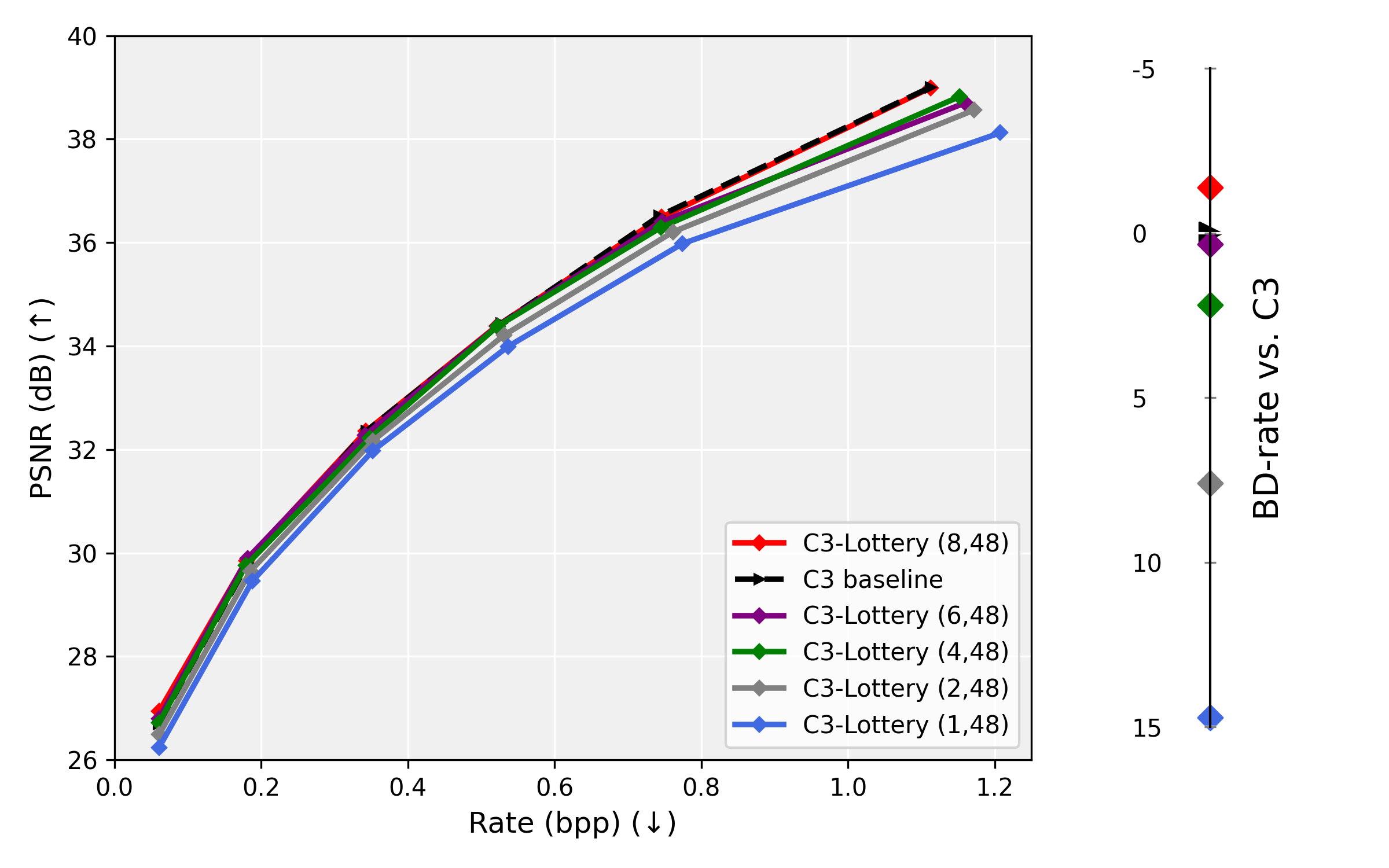}
        \label{op_psnr_rate_48}
    }%
    \subfloat[]{
        \centering
        \includegraphics[width=0.47\linewidth]{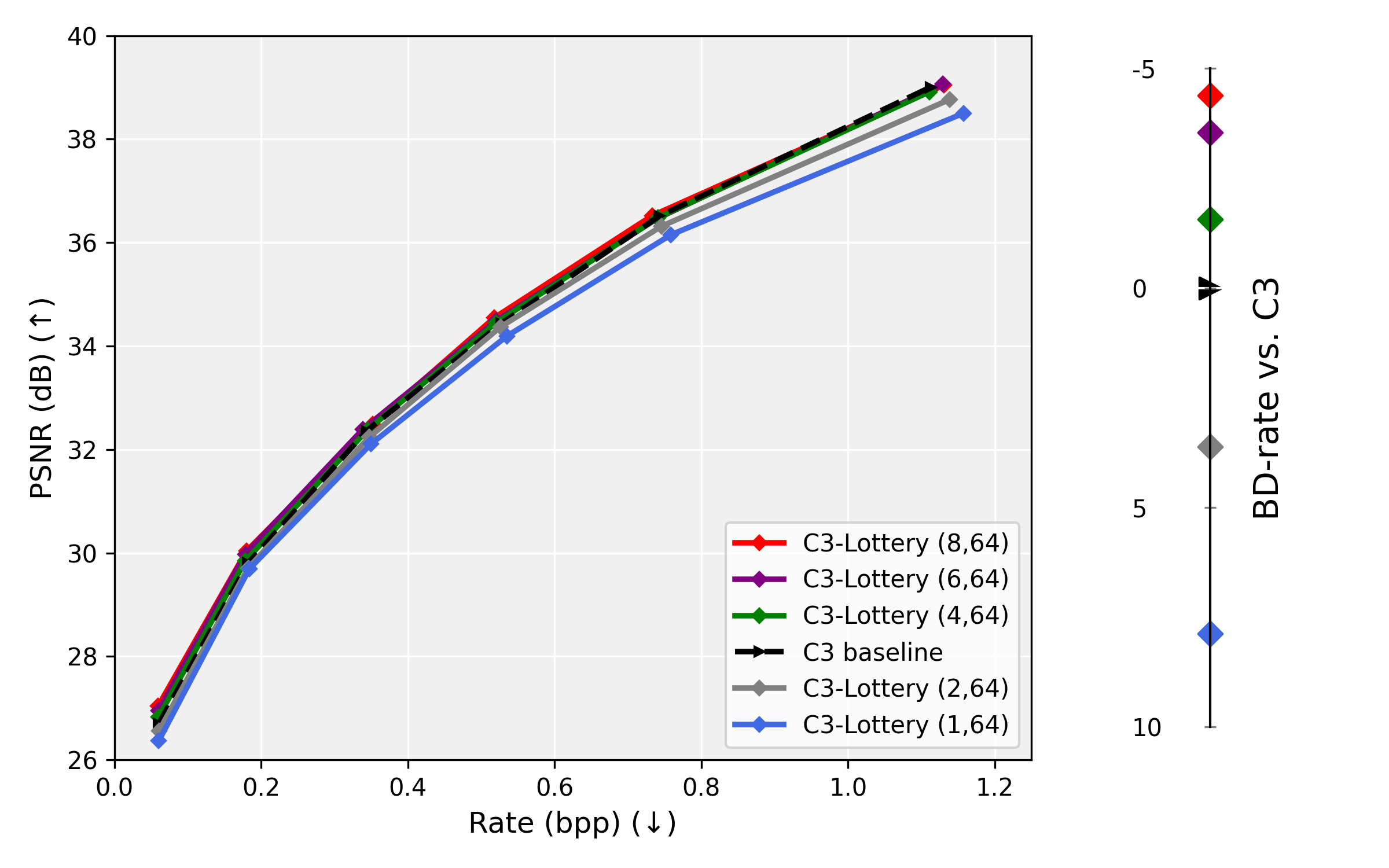}
        \label{op_psnr_rate_64}
    }%
    \caption{Verification of the \textit{lottery codec hypothesis} across varying network depths when hidden dimension is $48$ and $64$.}
    \label{op_psnr_rate_48_64}
\end{figure*}

\paragraph{More experimental results.} 
We also provide additional experimental results on the impact of increasing the depth of the over-parameterized network, as shown in Fig. \ref{op_psnr_rate_48_64}. Specifically, Fig. \ref{op_psnr_rate_48} and Fig. \ref{op_psnr_rate_64} present the PSNR vs. rate term of $\hat{\mathbf{z}}$ (bpp) and the corresponding BD-rate performance for randomly initialized networks with hidden dimensions of $48$ and $64$, respectively. We observe that, in both cases, a deeper network generally increases the likelihood of finding a winning lottery ticket, leading to improved RD performance. Notably, for $d=64$, lower distortion is achieved with fewer layers. 



\begin{table}[!h]
\centering
\begin{tabular}{ccccc}
\hline
\textbf{C3-Lottery model} & {C3-Lottery $(4,24)$} & {C3-Lottery $(4,32)$} & {C3-Lottery $(4,64)$} & {C3-Lottery $(4,128)$} \\ \hline
{Masked linear layers} & $(7,24)$ &$(7,32)$ & $(7,64)$ &$(7,128)$  \\
                      &  $(24,24) \times 4$ &  $(32,32) \times 4$& $(64,64) \times 4$,& $(128,128) \times 4$  \\
                      & $(24,3)$ &$(32,3)$ & $(64,3)$ &$(128,3)$\\ \hline
\textbf{Baseline C3 } & \multicolumn{4}{c}{$(7,18); (18,3); (3,3); (3,3)$} \\ \hline
\end{tabular}
\caption{Going wider: Model architectures (input-output channels) for over-parameterization experiments, where all models employ an auto-regressive entropy model (ARM-$16$). The mask ratios for the experiments are $\{0.1, 0.3, 0.5, 0.7, 0.9$\}.}
\label{table:conv_models_go_wider}
\end{table}

\begin{table}[!h]
\centering
\begin{tabular}{cccccc}
\hline
\textbf{C3-Lottery model}  & C3-Lottery (1,48) & C3-Lottery (2,48) & C3-Lottery (4,48)  & C3-Lottery (6,48) &C3-Lottery (8,48) \\ \hline
{Masked linear layers}  &$(7,48)$ &$(7,48)$ & $(7,48)$& $(7,48)$ & $(7,48)$  \\
                      &  $(48,48) \times 1$ &  $(48,48) \times 2$& $(48,48) \times 4$, & $(48,48) \times 6$ & $(48,48) \times 8$  \\
                    &$(48,3)$& $(48,3)$ & $(48,3)$ &$(48,3)$ &$(48,3)$\\ \hline
\textbf{C3-Lottery model}  & \textbf{$O_{64-1}$} & \textbf{$O_{64-2}$} & \textbf{$O_{64-4}$}  & \textbf{$O_{64-6}$}& \textbf{$O_{64-8}$}  \\ \hline
{Masked linear layers}  &$(7,48)$ &$(7,64)$ & $(7,64)$& $(7,64)$  & $(7,64)$ \\
                      &  $(64,64) \times 1$ &  $(64,64) \times 2$& $(64,64) \times 4$, & $(64,64) \times 6$ & $(64,64) \times 8$  \\
                    &$(64,3)$& $(64,3)$ & $(64,3)$ &$(64,3)$ &$(64,3)$\\ \hline
\textbf{Baseline C3 } & \multicolumn{5}{c}{$(7,18); (18,3); (3,3); (3,3)$} \\ \hline
\end{tabular}
\caption{Going deeper: Model architectures (input-output channels) for over-parameterization experiments, where all models employ an auto-regressive entropy model (ARM-$16$). The mask ratios for the experiments are $\{0.3, 0.5, 0.7$\}.}
\label{table:conv_models_go_deeper}
\end{table}

\begin{figure*}[t]
    \centering
    \subfloat[Kodak]{
                \centering
\includegraphics[width=0.45\linewidth]{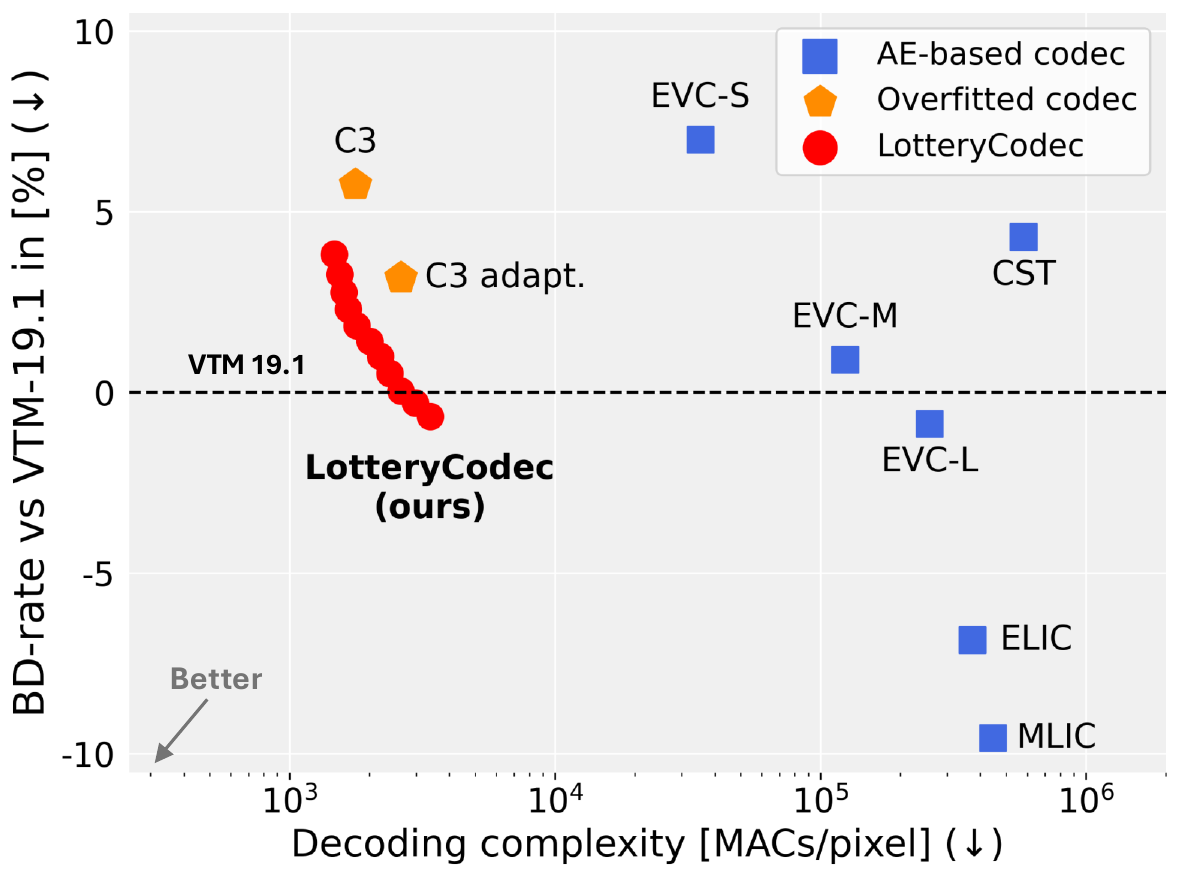}
        \label{bd_copmlexity_kodak}
    }%
    \subfloat[CLIC2020]{
            \centering
    \includegraphics[width=0.45\linewidth]{image/bd_vs_complex_clic.pdf}
        \label{bd_copmlexity_clic}
    }%
   \caption{Performance of LotteryCodec (mask ratios $\in[0.15:0.05:0.9]$) across different decoding complexities (a) BD-rate across different decoding complexities on Kodak dataset. (b) BD-rate across different decoding complexities on the CLIC2020 dataset.}
    \label{bd_vs_complexity}
\end{figure*}

\subsection{\textcolor{black}{More experiments: BD rate vs. decoding complexity}}
\label{ap_bd_complexity}
We report the BD-rate (vs. VTM 19.1) on the Kodak and CLIC dataset in Fig. \ref{bd_vs_complexity}. We note that VTM configurations vary between implementations~\cite{kim2024c3,blard2024overfitted}, and the BD-rate computation depends on both configurations and datapoints. For a fair and more aligned comparison, we update VTM baseline into VTM-19.1 from CompressAI~\cite{begaint2020compressai} and recompute BD-rates for all codecs under similar $\lambda$ settings. We also open-resourced all above baselines and datapoints in our project page for future alignment.

Due to computational constraints, the optimal BD-rate is evaluated over ratio $[0.15, 0.9]$ and $\lambda\in\{1e^{-2},5e^{-3},1e^{-3},5e^{-4},2e^{-4},1e^{-4}\}$. Other datapoints in Fig. \ref{main_result} (especially CLIC2020) are trained with a narrower range $[0.15, 0.45]$, yet LotteryCodec still achieves strong performance. Further optimized RD–complexity trade-offs are expected with broader mask ratio selection (see our updates on the project page). Detailed datapoints, lambda, and complexity are provided on our project page.



\subsection{\textcolor{black}{Theoretical  vs. practical decoding complexity}}
The current figure reports the theoretical minimum decoding complexity, excluding the multiplication operations of masked parameters, an evaluation approach also adopted in \cite{han2015deep,han2015learning}. This lower bound can be approached with sparsity-aware libraries (e.g., TVM \cite{chen2018tvm}, cuSparse \cite{naumov2010cusparse}, DeepSparse \cite{pmlr-v119-kurtz20a}) on appropriate hardware. We adopt this metric because practical MACs per pixel and runtime for unstructured sparse networks are highly dependent on implementation-specific engineering factors, making fair comparisons difficult to conduct.

\begin{figure*}[t]
    \centering
    \subfloat[Kodak]{
                \centering
\includegraphics[width=0.45\linewidth]{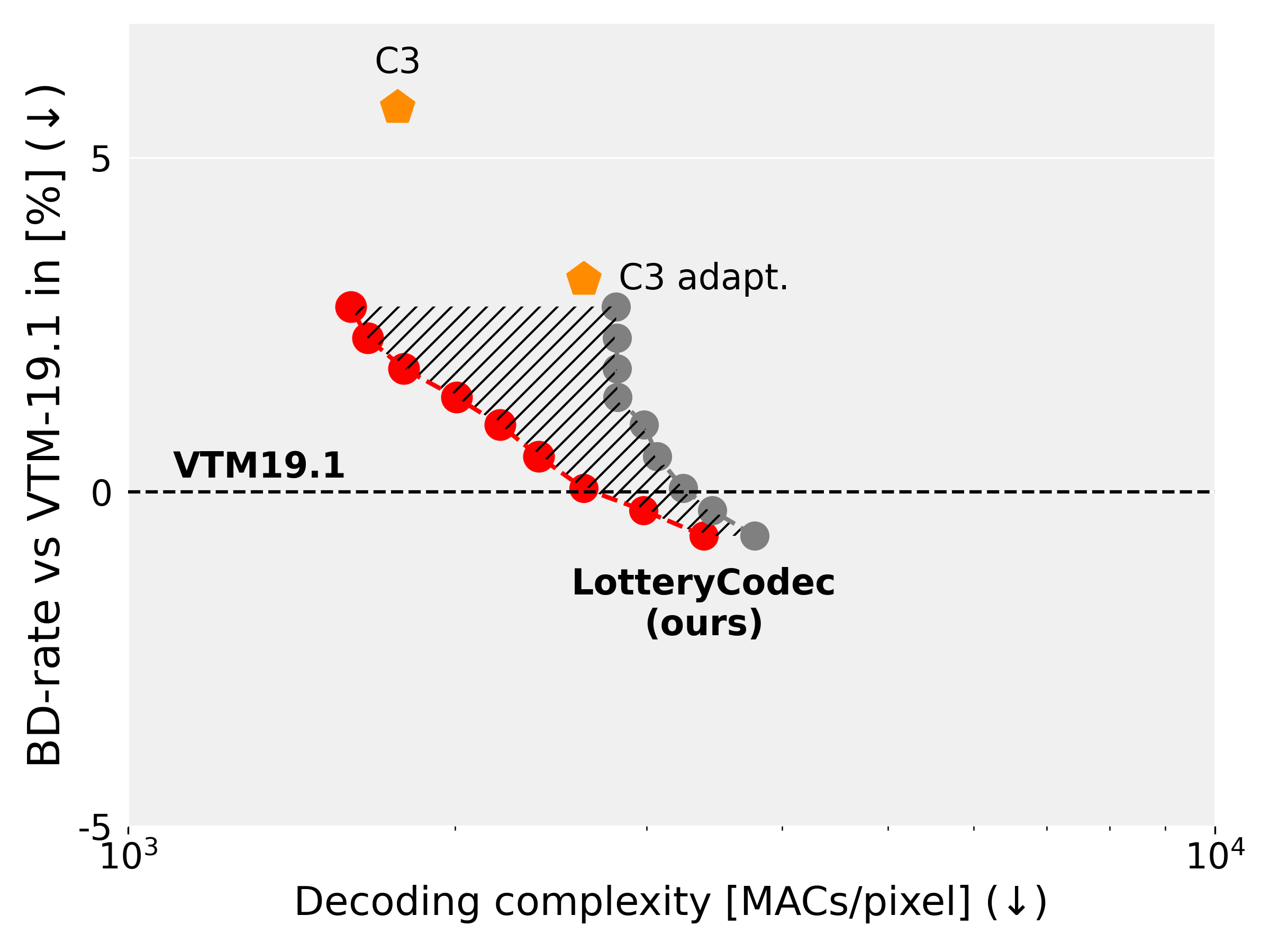}
    }%
    \subfloat[CLIC2020]{
            \centering
    \includegraphics[width=0.45\linewidth]{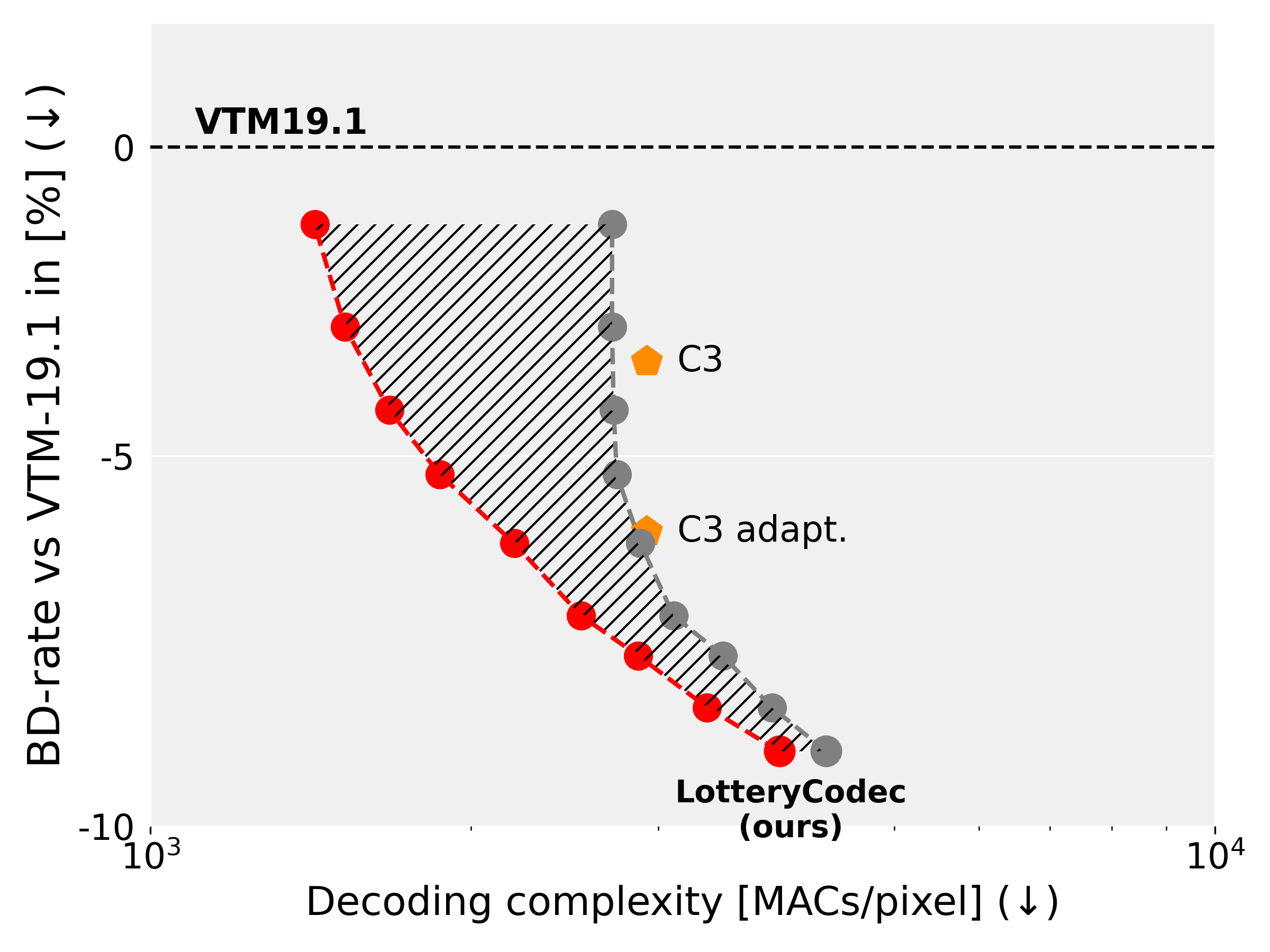}
    }%
    \caption{Flexible complexity region for Kodak and CLIC2020, where the dashed region is achievable via varying the mask ratios.}
    \label{fig:ab_flexible_BD}
\end{figure*}

For a comprehensive analysis, we report real coding times using a simple structured pruning strategy (see Tables \ref{tab:encoding_time} \ref{tab:encoding_time_across_resolution}), highlighting LotteryCodec’s efficiency, especially on high-resolution images. Additionally, we also provide both theoretical upper and lower bounds on decoding complexity: the upper bound accounts for all operations without pruning, while the lower bound considers only active components. Practical complexity lies between these bounds, depending on implementation. {As shown in Fig. \ref{fig:ab_flexible_BD}, even without pruning, LotteryCodec achieves better BD-rate than C3 scheme, with comparable complexity.}

\begin{figure*}[t]
    \centering
    \subfloat[]{
        \includegraphics[width=0.43\linewidth]{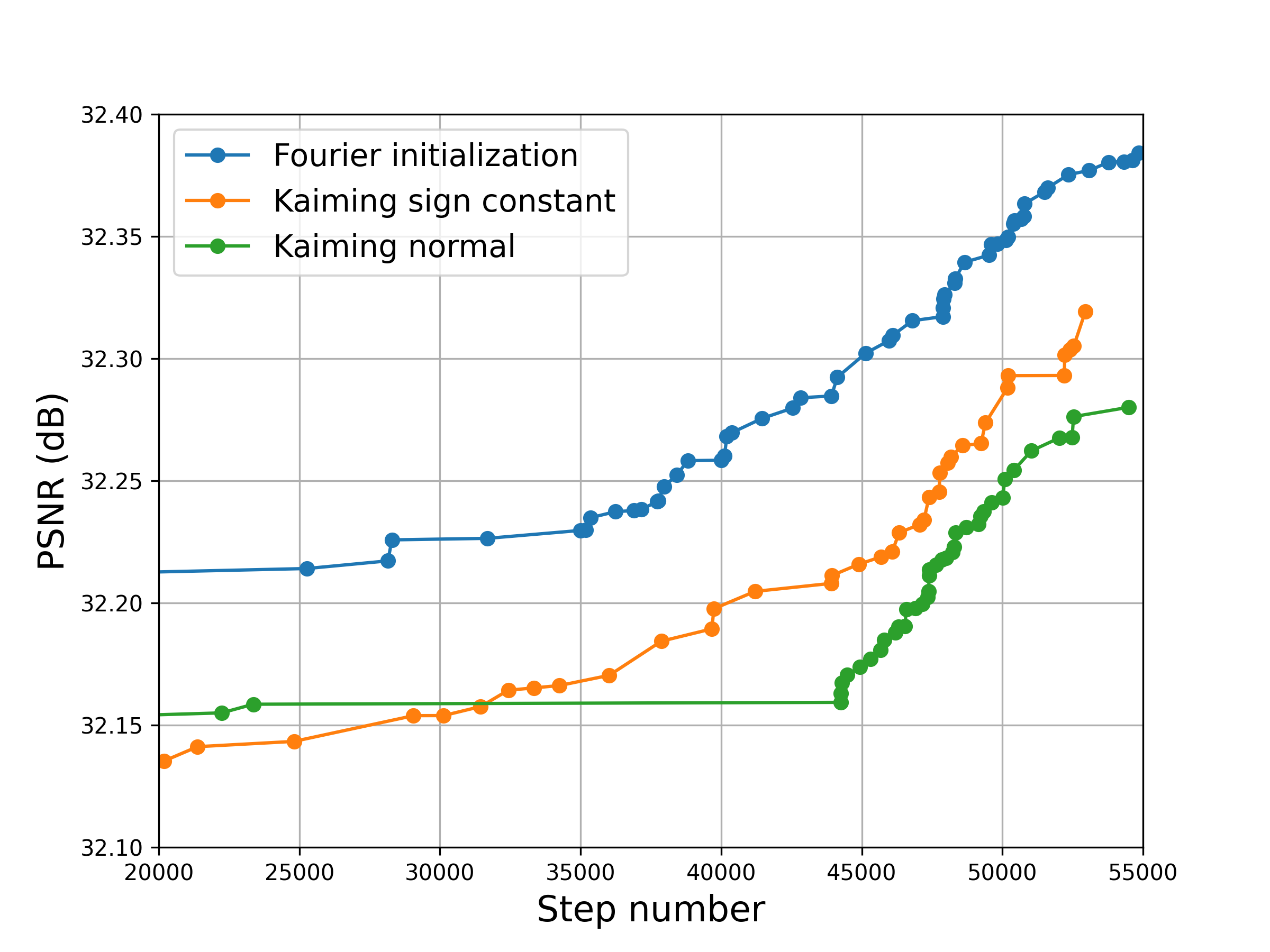}
        \label{ab_init_psnr}
    }%
        \hspace{+2pt}
    \subfloat[]{
        \includegraphics[width=0.43\linewidth]{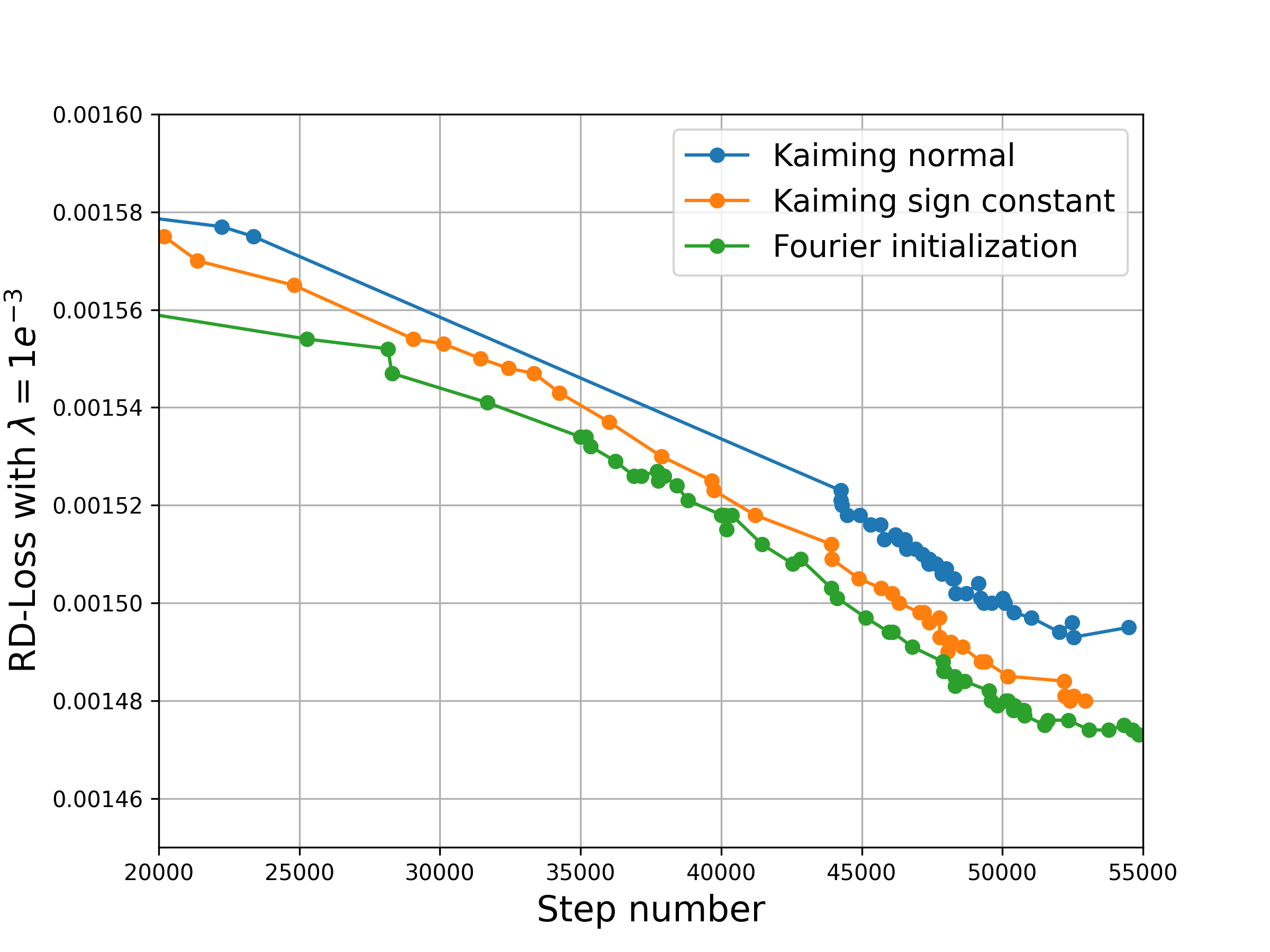}
        \label{ab_init_rate}
    }%
   \caption{Ablation study on initialization methods: (a) PSNR performance vs. coding step. (b) RD performance vs. coding step. The model is evaluated every $10$ coding steps, and the best-performing model at each step is plotted. Results are presented for kodim01 as an example.}
    \label{ab_init}
\end{figure*}

\section{\textcolor{black}{Qualitative analysis}}
\paragraph{Justification of Lottery Codec Hypothesis (LCH).} Although a rigorous bound supporting the LCH is not available, we can provide a rough validation based on existing proofs for Strong Lottery Tickets Hypothesis (SLTH). Suppose a codec $g_{\mathbf{{W}}}(\mathbf{z})$ is overfitted to an image $\mathbf{S}$, resulting in distortion $\sigma$. According to SLTH, for any $\epsilon>0$, there exists a subnetwork within a sufficiently over-parameterized network $g_{\mathbf{W'}}$, defined by a supermask $\tau$, such that $d(g_{\mathbf{W}}(\mathbf{z}),g_{{\mathbf{W'}}\odot \tau}({\mathbf{z}})) \le \epsilon$. Thus, reconstructing the image $\mathbf{S}$ using $g_{\mathbf{W'}\odot \mathbf{\tau}}(\mathbf{z})$ results in a distortion of at most $\sigma+\epsilon$. Now, we can further decrease the distortion by optimizing the latent vector over a set of $\mathbf{z'}$ satisfying $H(\mathbf{z'})=H(\mathbf{z})$, along with the supermask $\tau$. Since $\epsilon$ can be made arbitrarily small, it is highly likely that we can find a pair of $(\mathbf{\tau'},\mathbf{z'})$ such that $d(\mathbf{S},g_{\mathbf{W'}\odot \mathbf{\tau'}}({\mathbf{z'}}))\le \sigma$.

\paragraph{Intuition Behind the Advantages of LotteryCodec.} Based on the LCH, we can intuitively justify why the proposed LotteryCodec outperforms previous overfitted codecs in terms of the rate-distortion performance. The rate formulations for overfitted codecs and our LotteryCodec are given in Eqs. (2) and (5), respectively. They show that the rate of overfitted codecs depends on $\{\mathbf{\hat{z}}, \mathbf{\hat{\psi}}, \mathbf{\hat{W}}\}$, while our method is determined by $\{\mathbf{\hat{z}}, \mathbf{\hat{\psi}}, \mathbf{\tau}, \mathbf{\hat{\theta}}\}$. According to LCH, to achieve the same level of distortion, we can find a pair of $(\mathbf{\hat{z}},\mathbf{\tau})$ such that the bit cost for $\mathbf{\hat{z}}$ and $\mathbf{\hat{\psi}}$ is equal to that of overfitted codecs. While each quantized parameter in $\mathbf{\hat{W}}$ typically requires over 13 bits, our binary mask $\mathbf{\tau}$ uses just 1 bit per entry. Despite its higher dimensionality, $\mathbf{\tau}$ contributes significantly less to the total rate. Moreover, since $\mathbf{\hat{\theta}}$ is lightweight, the combined rate of $\mathbf{\tau}$ and $\mathbf{\hat{\theta}}$ remains lower than that of $\mathbf{\hat{W}}$, resulting in a lower compression rate and improved RD performance.

\color{black}
\section{Ablation studies}
\subsection{\textcolor{black}{Impact of each component in LotteryCodec}}
To clarify the impact of each component in our design, we conduct an ablation study on each component. As shown in Table \ref{tab:ablation_each} below, removing the Supermask network and using only the modulation network increases BD-rate by $+12.45\%$, highlighting the importance of the random network. Removing ModNet and directly feeding $\mathbf{z}$ into the random network (with different overparameterization configurations) results in a performance drop of up to $+14.99\%$ due to high overparameterization costs. Additional ablation studies and visualizations of other components are provided in Table \ref{table:ablation_study}.

\begin{table}[h]
    \centering
    \begin{tabular}{|c|c|c|c|}
    \hline
    \textbf{LotteryCodec} & \textbf{w/o SuperMask} & Random network \textbf{w/o ModNet}: $(4,32)/(4,48)/(4,64)$ \\
    \hline
    0 &  $+12.45\%$ & $+13.02\% / +11.98\% / +14.99\%$  \\
    \hline
    \end{tabular}
    \caption{Change in BD-rate due to removal of individual components from LotteryCodec, ARM dimension $16$.}
    \label{tab:ablation_each}
    \end{table}
\subsection{\textcolor{black}{MS-SSIM distortion metric}}
\begin{figure}[h]
    \centering
\includegraphics[width=0.55\linewidth]{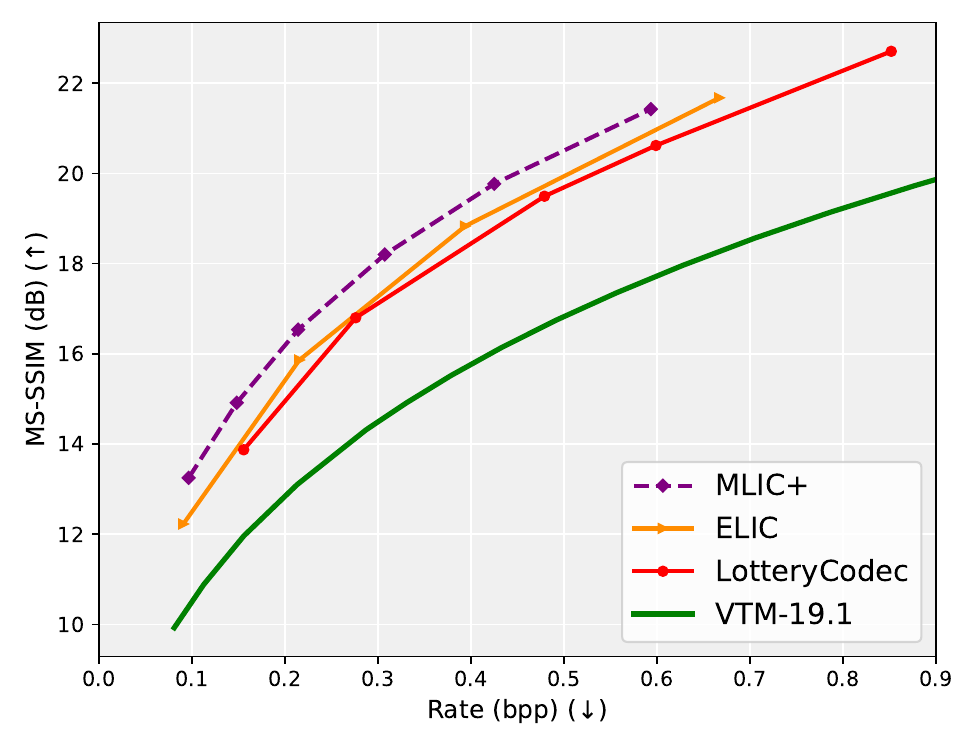}
    \caption{Rate-distortion performance for MS-SSIM metric.}
    \label{fig:MS_SSIM}
\end{figure}

In addition, we evaluate LotteryCodec against baseline methods in terms of MS-SSIM distortion on the Kodak dataset, as shown in Fig.~\ref{fig:MS_SSIM}. Specifically, we train our method with an MS-SSIM loss with $\lambda\in\{3e^{-1},1e^{-1},3e^{-2},2e^{-2}, 1e^{-2}\}$. It achieves up to a $-43.39\%$ BD-rate reduction over VTM-19.1, closely approaching ELIC and showing a $+10.41\%$ BD-rate gap compared to MLIC+. 

\subsection{Other initialization methods}
We conduct different initialization methods, the overall BD performance is presented in Table \ref{table:ablation_study}. To simplify the validation, here we employ the same training steps of $60000$ in Stage I and $6000$ for Stage II for all schemes. We also present the PSNR performance and loss values during the coding process in Fig. \ref{ab_init}. We can observe that the Fourier initialization we employed can enhance the performance and coding process.

\begin{table*}[h]
    \centering
    \renewcommand\arraystretch{1.6} 
    \setlength{\tabcolsep}{5mm} 
    \resizebox{0.7\textwidth}{!}{ 
    \begin{tabular}{l c}
    \toprule
    \textbf{Model Variant} & \textbf{BD rate vs. \textit{LotteryCodec}}  \\
    \midrule
   \textit{LotteryCodec} scheme & {$0.0\%$}\\
     \quad $\Rightarrow$ Kaiming normal initialization \cite{he2016deep}&  \textcolor{orange}{$ +1.55\%$}\\
   \quad $\Rightarrow$ Signed Kaiming constant initialization \cite{ramanujan2020s} &  
   \textcolor{orange}{$+0.63\%$}\\
  \quad $\Rightarrow$ FilM modulation methods \cite{perez2018film} &  \textcolor{orange}{$ +2.21\%$}\\
   \quad $\Rightarrow$ Score-based same layer masking algorithm \cite{ramanujan2020s} &  \textcolor{orange}{$+1.77\%$}\\
     \quad $\Rightarrow$ NeRF positional encoding module \cite{mildenhall2021nerf} &  \textbf{Not work well} \\ 
   \quad $\Rightarrow$ Xavier normal initialization \cite{he2016deep} &  \textbf{Not work well} \\ 
      \quad $\Rightarrow$ Gumble-softmax for mask ratio learning \cite{milescascaded,9897718}&  \textbf{Not work well} \\ 
   \quad $\Rightarrow$ Bernoulli-based masking algorithm \cite{zhou2019deconstructing} &  \textbf{Not work well} \\ 
    \bottomrule
    \end{tabular}
    }
    \caption{{Ablation study on different mechanisms, conducted on the first $10$ Kodak images using an ARM-$16$ model and a mask ratio of $20\%$, where $\lambda\in\{2e^{-2},1e^{-2},5e^{-3},1e^{-3},2e^{-4},\}$. A higher BD rate indicates worse RD performance, and ``Not work well'' signifies lack of robustness and unsatisfactory results}}
    \label{table:ablation_study}
\end{table*}

\subsection{Other Modulation methods}
\begin{figure}[h]
    \centering
\includegraphics[width=0.85\linewidth]{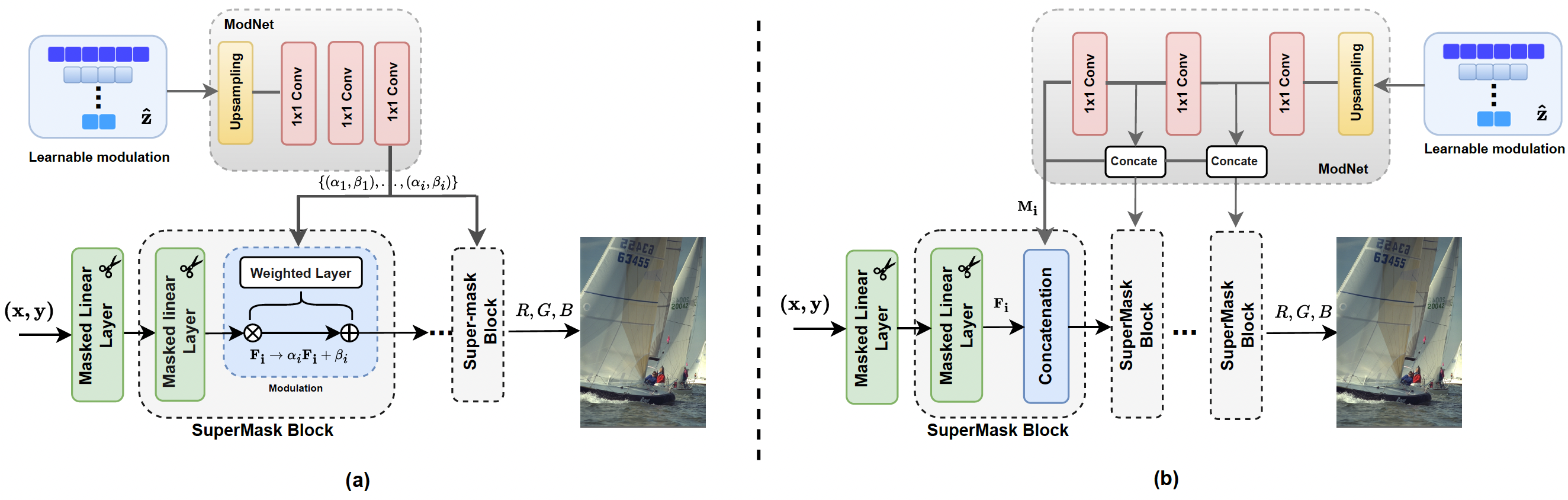}
    \caption{Different alternative modulation methods of the LotteryCodec. (a). A FilM-based modulation approach. (b). Concatenation-based approach}
    \label{fig:new_mod}
\end{figure}
\label{ap_mod}
Note that LotteryCodec is a flexible framework that supports various modulation methods. We also implement a FiLM-based modulation \cite{perez2018film}, as shown in Fig. \ref{fig:new_mod} (a), whose performance is presented in the Table \ref{table:ablation_study}. Interestingly, experiments show that additive bias alone achieves competitive performance, although it falls short of the proposed concatenation-based method. This approach, however, provides the potential of reducing inference complexity and exploring alternative modulation strategies in future research.
\subsection{Alternative masking learning approach}
\label{ap_mask}
This section presents various trials we explored. While some did not yield satisfactory results, we include them for completeness and future reference. Specifically, to achieve an adaptive mask ratio, we employed a Gumbel-Softmax scheme for dynamically controlling the pruning rate, following \cite{milescascaded,9897718}, but it did not produce satisfactory results. A similar Bernoulli-based approach, as in \cite{zhou2019deconstructing}, also underperformed compared to the current LotteryCodec. Additionally, we evaluated a mask learning algorithm applying a fixed pruning rate across each layer \cite{ramanujan2020s}, which resulted in worse performance. Lastly, positional encoding, similar to \cite{shi2024improved}, did not yield performance gains in the context of overfitted codecs.

\subsection{\textcolor{black}{Coding latency across various resolutions}}
We present the coding cost of various schemes in Table \ref{tab:encoding_time}, and report resolution-dependent coding costs in Table \ref{tab:encoding_time_across_resolution}. The proposed method shows scalability to ultra-high-resolution images, albeit with increased coding time. An additional example of 2K image encoding is shown in Table \ref{tab:ablation_steps}. Overall, our method has a slightly higher encoding time than other overfitted codecs due to additional gradient-based mask learning, but it offers greater flexibility and faster decoding. Notably, the lottery codec hypothesis provides potential for parallel encoding by re-parameterizing distinct network optimizations into batch-wise mask learning, highlighting its advantage of scalability for efficient large-scale image encoding.

All of these results are based on unoptimized code and current hardware, which can be significantly improved with proper engineering optimization. For example, we can accelerate inference via ONNX and DeepSparse libraries to reduce the decoding time to $20–80$ ms on a CPU. Additional techniques \cite{blard2024overfitted}, such as symmetric/separable kernels, filter-based upsampling, and wavefront decoding, can further enhance the speed of overfitted codecs. 

 \begin{table}[h]
    \centering
    \resizebox{0.8\textwidth}{!}{
    \begin{tabular}{|c|c|c|c|}
    \hline  
    {Models} & { {Encoding time} }& {Decoding time } \\
       \hline
    { {VTM 19.1}} & {{\textcolor{blue}{$85.53$ (s)} }}& {\textcolor{blue}{ $352.52$ (ms) }}\\
    \hline
   {EVC} (S/M/L) & \textcolor{orange}{$20.23/32.21/51.35$ (ms)} & \textcolor{orange}{$18.82/23.73/32.56$ (ms)} \\
    MLIC+ & \textcolor{orange}{$205.60$ (ms)} & \textcolor{orange}{$271.31$ (ms)}\\
\hline
    {LotteryCodec} ($d=8/16/24$) & \textcolor{orange}{ $13.86$/$14.64$/$14.92$ (sec/1k steps)} & \textcolor{blue}{$261.33/267.58/278.31$ (ms)}\\
    {C3 }($d=12/18/24$) & \textcolor{orange}{$13.10/13.98/14.32$ (sec/1k steps)}& \textcolor{blue}{$272.15/284.67/295.03$ (ms)} \\
    \hline
    \end{tabular}
    }
    \caption{Coding time for Kodak images on NVIDIA L40S (GPU) and Intel Xeon Platinum 8358 (CPU) with a masking ratio of $0.8$ under structured pruning. Orange indicates \textcolor{orange}{GPU computation}; blue indicates \textcolor{blue}{CPU computation}.}
    \label{tab:encoding_time}
    \end{table}

\begin{table}[h]
    \centering
    \begin{tabular}{|c|c|c|c|l|}
    \hline  
  &  \multicolumn{2}{|c|}{\textbf{LotteryCodec vs. C3 }} & {\textbf{LotteryCodec vs. C3 vs. MLIC+}}  \\
\hline
   \textbf {Input resolution} & {\textbf{GPU Encoding} }& \textbf{CPU Decoding }& \textbf{Peak Memory usuage during the training}\\
& {(sec/1k steps) }& (ms)& {(GB)}\\
    \hline
$512\times 512$ & $10.71$ vs. ${10.43}$ &  ${232.46}$ vs. $228.43$ &{${0.56}$} vs. ${0.31}$ vs. $1.98$  \\
\hline
$1024\times 1024$ & $56.81$ vs. $38.54$ & ${565.22}$ vs. $576.51$ & \textbf{${2.15}$} vs.  {${1.24}$} vs. $3.61$  \\
\hline
$1536\times 1536$ & $136.81$ vs. ${84.79}$ & ${984.01}$ vs. $1086.92$& \textbf{${4.82}$} vs. \textbf{${2.78}$} vs. $9.15$  \\
\hline
$2048\times 2048$ & $257.93$ vs. ${155.02}$  & ${1595.86}$ vs. $1807.35$& \textbf{${8.53}$} vs. \textbf{${4.95}$} vs. $24.37$  \\
\hline
$2560\times 2560$ & $407.68$ vs. ${237.45}$ & ${3003.24}$  vs. $3269.02$ & \textbf{${13.36}$} vs. \textbf{${7.72}$} vs. OM \\
\hline
$3840\times 2160$ & {$446.09$} vs. {${301.56}$} & {${4014.21}$}  vs. $4216.11$ & \textbf{${16.89}$} vs. \textbf{${9.84}$} vs. OM \\
\hline 
\end{tabular}
\caption{Encoding time for images across different resolutions, where mask ratio $0.8$ and ARM model of $d=16$ are employed. OM means out of memory ($>32$ GB).} 
    \label{tab:encoding_time_across_resolution}
\end{table}

\begin{table}[h]
    \centering
    \begin{tabular}{|c|c|c|c|c|}
    \hline
    \textbf{Training steps} & \textbf{Training time (s)} & \textbf{bpp} & \textbf{PSNR (dB)} \\
    \hline
    5k & 678 & $0.24$ & $36.51$  \\
    10k & 1347 & $0.22$ & $36.92$  \\
    20k & 2685 &  $0.21$ & $37.02$  \\
    30k & 4026 & $0.20$ & $37.10$  \\
    50k & 6733&  $0.199$ & $37.14$  \\
    \hline
    \end{tabular}
    \caption{Encoding cost for a 2K image (size $1292 \times 1945$), ``davide-ragusa-716 in CLIC2020 with optimal result PSNR $37.18$ at bpp  $0.196$'' ($d=24$, ratio $0.2$, peak memory 5.64 G), where 10-20k steps can yield a descent performance.}
    \label{tab:ablation_steps}
\end{table}

\section{Pseudocode for the algorithm}
\label{pse_algorithm}
This section provides detailed encoding and decoding algorithm of LotteryCodec, as shown in Algorithm \ref{alg_train} and  Algorithm \ref{alg_infer}.

\section{Visualization}
\label{ap_vis}
\begin{figure*}[!h]
    \centering
    \subfloat[]{
        \includegraphics[width=0.4\linewidth]{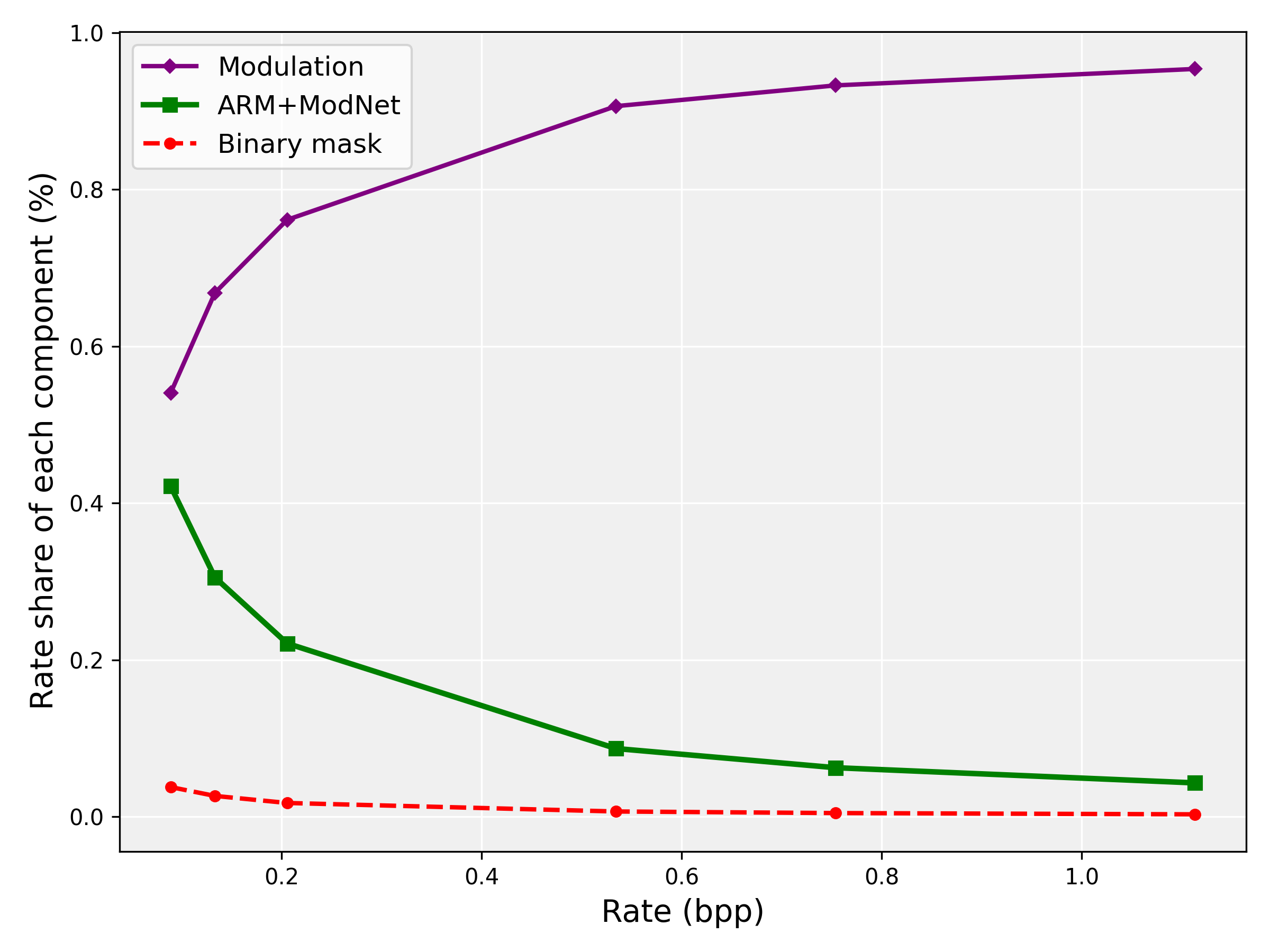}
        \label{cost_kodak}
    }%
        \hspace{+2pt}
    \subfloat[]{
        \includegraphics[width=0.4\linewidth]{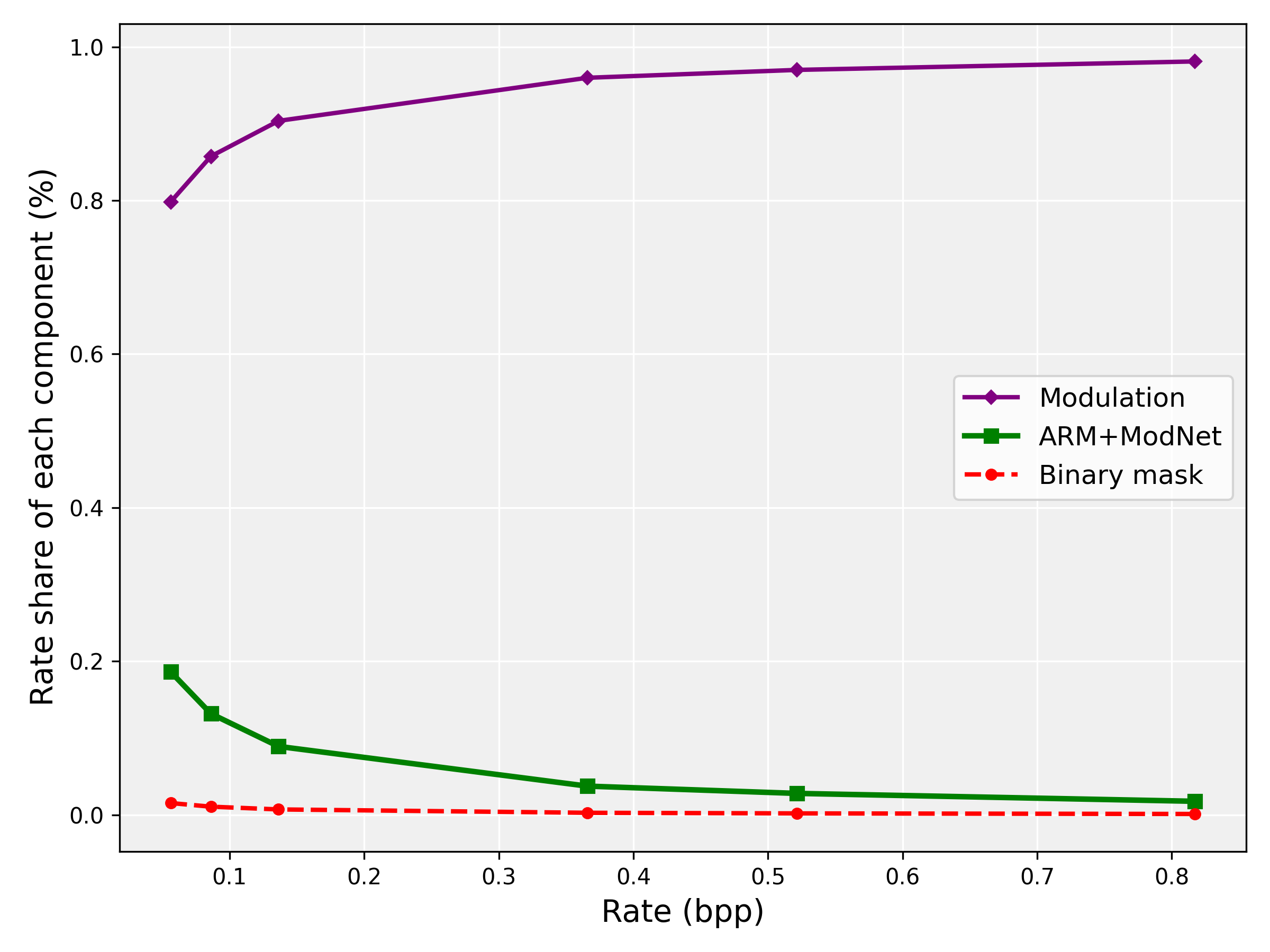}
        \label{cost_clic}
    }%
   \caption{Visualization of the compression cost distribution across the rate within the LotteryCodec scheme using a ARM-$24$ model and a mask ratio of $0.2$: (a) Rate share of compression cost on the Kodak. (b) Rate share of compression cost on the CLIC2020.}
    \label{cost_compression}
\end{figure*}

\paragraph{Compression Cost.} 
This section visualizes the compression cost of each component in the bitstream of LotteryCodec. As shown in Fig. \ref{cost_compression}, thanks to the introduction of ModNet, bit cost of the binary mask can be minimal, particularly for higher-resolution images like those in the CLIC2020 dataset, where the same network is used with a lower bpp contribution from binary mask. As bpp increases, modulation bit cost rises, allowing finer image details to be preserved within the given bit budget. Since the network architecture is fixed, then the relative bit cost ($\%$) contribution from the network and binary mask decreases as bpp increases.

\begin{figure*}[t]
    \centering
    \includegraphics[width=0.4\linewidth]{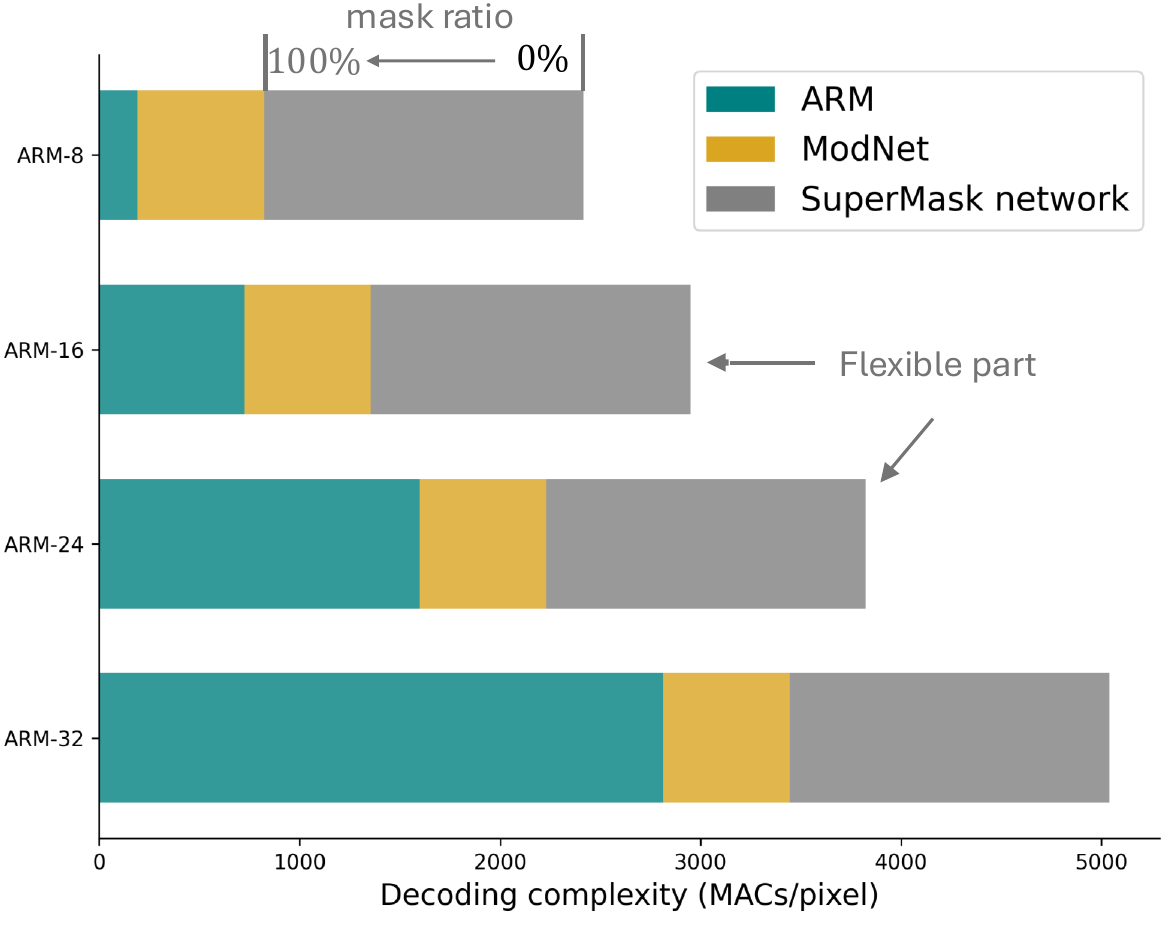}
    \caption{Decoding complexity of each component in LotteryCodec, where ARM-d denotes ARM model with a hidden dimension of $d$.}
\label{fig:complexity_component}
\end{figure*}

\begin{algorithm}[t] 
    \caption{Encoding stage of the {LotteryCodec}}
    \textbf{Input:} Source image $\mathbf{S}$, coordinate vector $\mathbf{x}$, random seed $e$,\\
    learning rates $\alpha$ for mask scores ($\mathbf{P}$), $\beta$ for modulations ($\mathbf{z}$), and networks (ModNet $\mathbf{\theta}$and ARM $\mathbf{\psi}$).\\
    cosine scheduler for learning rate for Stage I and II: $\mathcal{C}_1$, $\mathcal{C}_2$,\\
    linear scheduler for soft-rounding temperature and  Kumaraswamy noise strength for Stage I and II: ${l}_1$, ${l}_2$.\\
    Greedy search for quantization step: $\mathcal{G}$.\\
    \textbf{Output:} Bits stream of $\mathbf{{z}}$, $\mathbf{{\psi}}$, $\mathbf{{\theta}}$, $\mathbf{\tau}$: $\mathbf{b_{z}}$, $\mathbf{b_{\psi}}$, $\mathbf{b_{\theta}}$, $\mathbf{b_{\tau}}$.\\
    \vspace{-10pt}
    \begin{algorithmic}[1]
                 \STATE   $\mathbf{W_0}=\mathbf{\Lambda}\mathbf{B}$,
        \COMMENT{Fourier initialization for $g(\cdot)$ based on $e$}
                 \STATE   $\mathbf{P}\sim \mathcal{U}_{k}$,
        \COMMENT{Kaiming uniform initialization}
        \FOR{the $i$-th step within the {Stage I}}
          \STATE   $\mathbf{\tau}=h(\mathbf{P})$, $g =g_{\mathbf{\tau}\odot \mathbf{w_0}} $
        \COMMENT{Configure network by masking $k\%$ weights based on $\mathbf{P}$}\\
        \STATE $\mathbf{\hat{z}}=\mathcal{S}_{T}(\mathbf{z})+\mathbf{u}_{kum}$,   \COMMENT{Quantization-aware training}
                  \STATE   $\mathbf{\hat{S}}=g_{\mathbf{\tau}\odot \mathbf{w_0}}(f_{\mathbf{{\theta}}} (\mathbf{\mathbf{\hat{z}}}),\mathbf{x})$,
                  \COMMENT{Modulate the reconstruction}
                  \STATE   $\mathcal{L}= D(\mathbf{S},\mathbf{\hat{S}})+\lambda R_{\mathbf{\psi}}(\mathbf{\hat{z}})$, 
        \COMMENT{Compute the RD cost loss function}\\
            \STATE     $\mathbf{{P}}\leftarrow\mathbf{P}-\alpha \nabla_{\mathbf{P}}\mathcal{L}$,
    \COMMENT{\textcolor{orange}{Update the scores for mask}}
        \STATE     $\mathbf{{z}}\leftarrow\mathbf{z}-\beta \nabla_{\mathbf{z}}\mathcal{L}$,
     \COMMENT{\textcolor{blue}{Update latent modulations}}\\
        \STATE     $\mathbf{{\theta}}\leftarrow\mathbf{\theta}-\beta \nabla_{\mathbf{\theta}}\mathcal{L}$,
        \COMMENT{\textcolor{blue}{Update ModNet}}\\
    \STATE     $\mathbf{{\psi}}\leftarrow\mathbf{\psi}-\beta \nabla_{\mathbf{\psi}}\mathcal{L}$,
    \COMMENT{\textcolor{blue}{Update ARM}}\\
\STATE      $\alpha=\mathcal{C}_1(\alpha,i)$, $\beta=\mathcal{C}_1(\beta,i)$                         \COMMENT{{Update learning rate}}\\
    \STATE    $T={l}_1(T,i)$, $u_{kum}={l}_1(u_{kum},i)$                        \COMMENT{{Update noise strength}}\\ 
            \ENDFOR
            \FOR{the $i$-th step within the {Stage II}}
          \STATE   $\mathbf{\tau}=h(\mathbf{P})$, $g =g_{\mathbf{\tau}\odot \mathbf{w_0}} $
        \COMMENT{Mask $k\%$ weights based on updated $\mathbf{P}$}\\
        \STATE $\mathbf{\hat{z}}=Q(\mathbf{z})$,   \COMMENT{Hard rounding}
                  \STATE   $\mathbf{\hat{S}}=g_{\mathbf{\tau}\odot \mathbf{w_0}}(f_{\mathbf{{\theta}}} (\mathbf{\mathbf{\hat{z}}}),\mathbf{x})$,
                  \COMMENT{Modulate the reconstruction}
                  \STATE   $\mathcal{L}= D(\mathbf{S},\mathbf{\hat{S}})+\lambda R_{\mathbf{\psi}}(\mathbf{\hat{z}})$, 
        \COMMENT{Compute the RD cost loss}\\
            \STATE     $\mathbf{{P}}\leftarrow\mathbf{P}-\alpha \nabla_{\mathbf{P}}\mathcal{L}$,
    \COMMENT{\textcolor{orange}{Update the scores for mask}}
                \STATE     $\mathbf{{z}}\leftarrow\mathbf{z}-\beta \nabla_{\mathbf{z}}\mathcal{L}$,
                 \COMMENT{\textcolor{blue}{Update latent modulations}}\\
                        \STATE     $\mathbf{{\theta}}\leftarrow\mathbf{\theta}-\beta \nabla_{\mathbf{\theta}}\mathcal{L}$,
                        \COMMENT{\textcolor{blue}{Update ModNet}}\\
                        \STATE     $\mathbf{{\psi}}\leftarrow\mathbf{\psi}-\beta \nabla_{\mathbf{\psi}}\mathcal{L}$,
                        \COMMENT{\textcolor{blue}{Update ARM}}\\
      \STATE      $\alpha=\mathcal{C}_2(\alpha,i)$, $\beta=\mathcal{C}_2(\beta,i)$                         \COMMENT{{Update learning rate}}\\
                \STATE    $T={l}_2(T,i)$, $u_{kum}={l}_2 (u_{kum},i)$                        \COMMENT{{Update noise strength}}\\
            \ENDFOR
     \STATE     $\mathbf{b_{{z}}}=\mathcal{A}(Q(\mathbf{z}))$
        \COMMENT{Bit stream of $\mathbf{z}$ after quantization and entropy coding}\\  
    \STATE  $\mathbf{b_{{\tau}}}=\mathcal{A}(\mathbf{\tau})$
    \COMMENT{Bit stream of $\mathbf{\tau}$ after entropy coding}\\   
    \STATE     $\Delta_{\theta}, \Delta_{\psi} =
    \mathcal{G}(\mathbf{\theta
    },\mathbf{\psi
    })$
                    \COMMENT{Search for a optimal quantization step for networks}\\   
        \STATE     $\mathbf{b_{{\theta}}}=\mathcal{A}(Q(\mathbf{\theta},\Delta_{\theta}))$
        \COMMENT{Bit stream of $\mathbf{\theta}$ after quantization and entropy coding}\\ \STATE     $\mathbf{b_{{\psi}}}=\mathcal{A}(Q(\mathbf{\psi},\Delta_{\psi}))$
        \COMMENT{Bit stream of $\mathbf{\psi}$ after quantization and entropy coding}\\   
    \end{algorithmic}
    \vspace{+3pt}
    \label{alg_train}
\end{algorithm}

\begin{algorithm}[t] 
    \caption{Decoding stage of the {LotteryCodec}}
 \textbf{Input:} Coordinate vector $\mathbf{x}$, random seed $e$, Bits stream: $\mathbf{b_{z}}$, $\mathbf{b_{\psi}}$, $\mathbf{b_{\theta}}$, $\mathbf{b_{\tau}}$.\\
\textbf{Output:} Reconstruction of image $\mathbf{\hat{S}}$.\\
    \vspace{-10pt}
    \begin{algorithmic}[1]
                 \STATE   $\mathbf{W_0}=\mathbf{\Lambda}\mathbf{B}$,
        \COMMENT{Fourier initialization for $g $ based on a given seed $e$. }
          \STATE   $\mathbf{\tau}=\mathcal{A}(\mathbf{b_{\tau}})$, $g =g_{\mathbf{\tau}\odot \mathbf{w_0}}$
        \COMMENT{\textcolor{orange}{Entropy decode mask and configure SuperMask network locally}}\\
         \STATE   $\mathbf{\hat{\psi}}=Q(\mathcal{A}(\mathbf{b_{\psi}}))$, $\mathbf{\hat{\theta}}=Q(\mathcal{A}(\mathbf{b_{\theta}}))$,
        \COMMENT{\textcolor{blue}{Entropy decode and de-quantize parameters $\hat{\psi}$ and $\hat{\theta}$}}\\

          \STATE   $\mathbf{\hat{z}} =\mathcal{A}({f_{\mathbf{\hat{\psi}}}},{\mathbf{b_z})}$,
    \COMMENT{\textcolor{blue}{Entropy decode $\mathbf{z}$}}
          \STATE   $\mathbf{\hat{S}}=g_{\mathbf{\tau}\odot \mathbf{w_0}}(f_{\mathbf{{\hat{\theta}}}} (\mathbf{\mathbf{\hat{z}}}),\mathbf{x})$,
        \COMMENT{\textcolor{DarkGreen}{Reconstruct the source image}}\\
    \end{algorithmic}
    \vspace{+3pt}
    \label{alg_infer}
\end{algorithm}

\paragraph{Decoding Complexity Analysis.} To illustrate the adaptive complexity advantage, we visualize the decoding complexity of each component in the LotteryCodec scheme in Fig. \ref{fig:complexity_component}. The SuperMask network contributes significantly to the total decoding complexity, which can be adaptively controlled via different mask ratios, enabling flexible complexity management across various regimes. Specifically, for LotteryCodec using ARM-8, the SuperMask accounts for more than $60\%$ of the total complexity, demonstrating its remarkable adaptability and efficiency, particularly in low-bpp regimes.

\begin{figure*}[t]
    \centering
    \subfloat[]{
        \centering
        \includegraphics[width=0.453\linewidth]{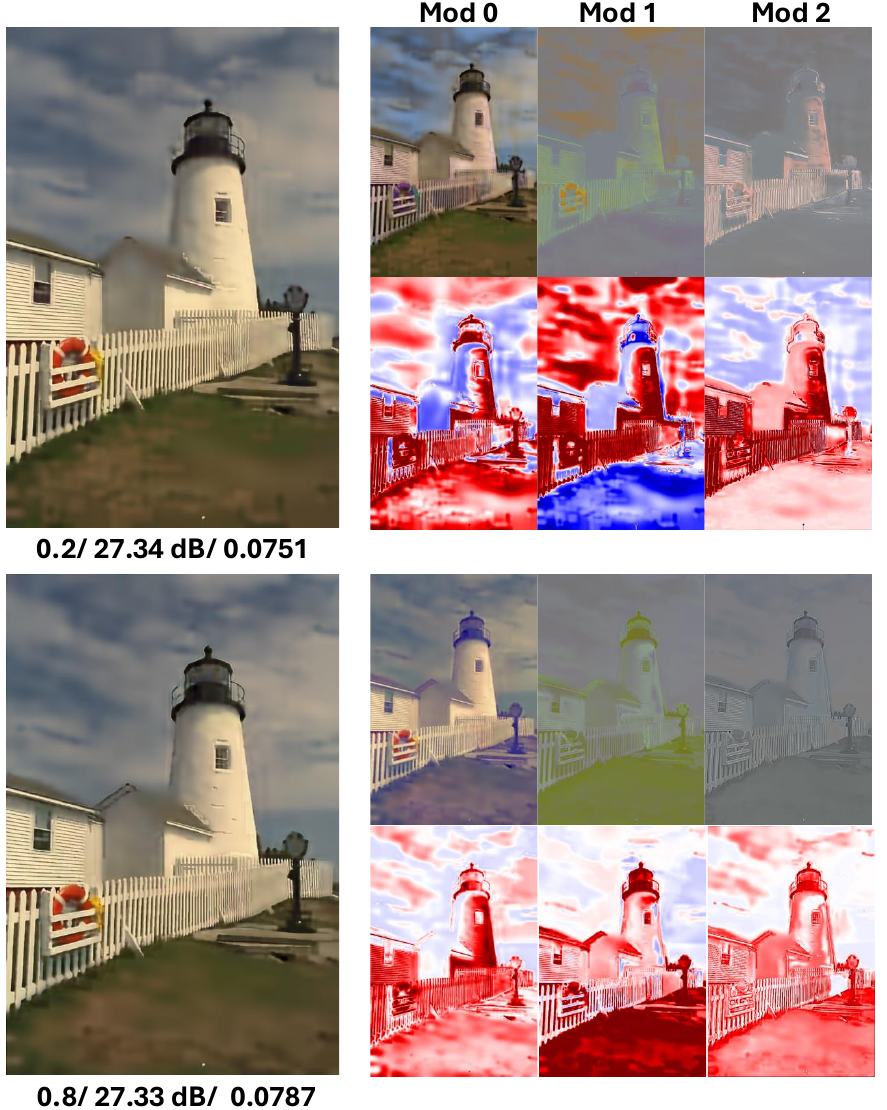}
    }
    \subfloat[]{
        \includegraphics[width=0.46\linewidth]{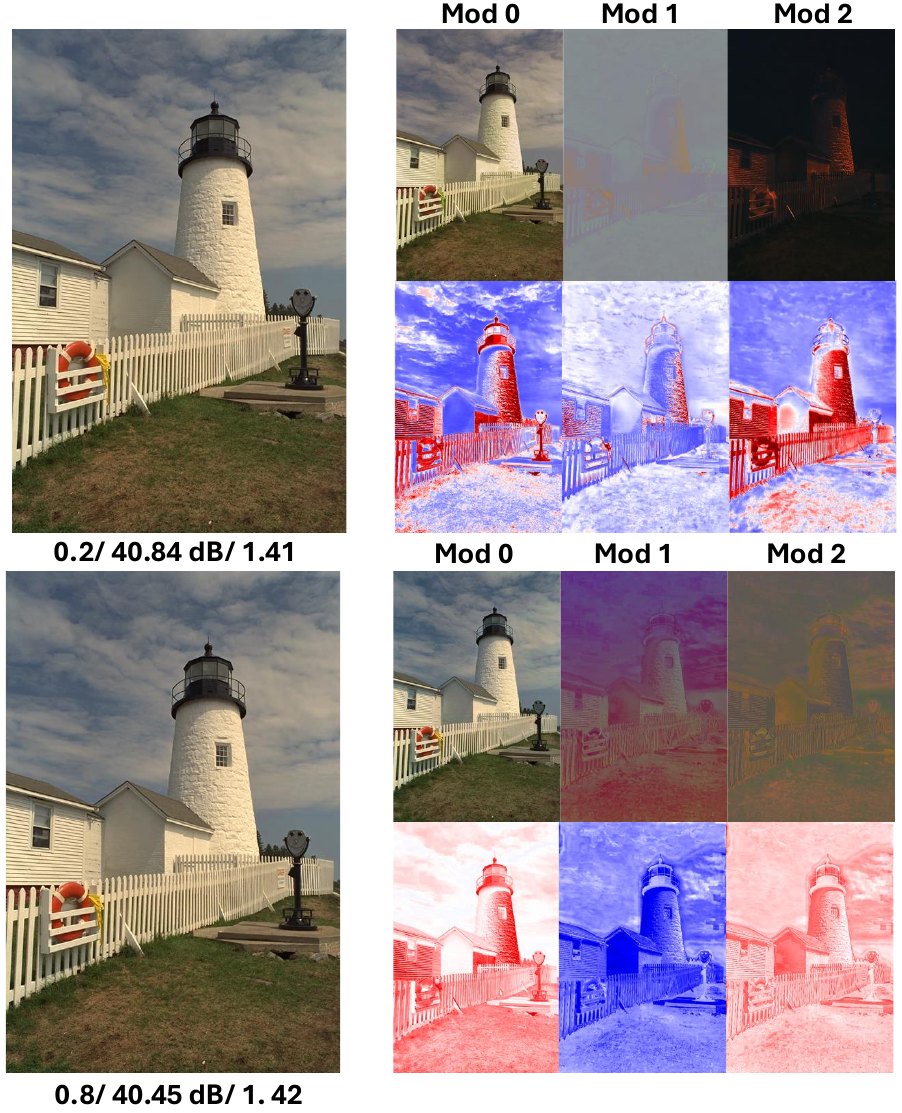}
    }%
    \caption{Visualization of ModNet outputs and their effects in LotteryCodec scheme, where $0.2/27.34/0.0751$ represent reconstruction with a $0.2$ mask ratio, $27.34$ PSNR, and $0.0751$ bpp. Mod $i$ represents the visualization of the $i$-th layer of ModNet. The first row shows reconstructions with all but the selected latent outputs set to zero, while the second row displays the averaged feature outputs, upscaled to match the resolution of the reconstructed image. Visualizations are provided for both low bit-rate (a) and high bit-rate (b) scenarios.}
    \label{vis_modulation_effect}
\end{figure*}

\paragraph{Effect of modulations in different SuperMask layers.} We visualize the effect of different modulations from each ModNet layer in Fig. \ref{vis_modulation_effect}, where we employ a three-layer ModNet and a four-layer SuperMask network. When considering the mask ratio: a lower mask ratio tends to require more complex modulations to aid the LotteryCodec to search the subnetwork, which is reflected in the increased variability of signs and entropy, leading to stronger representational ability. The variability of signs also indicates a greater dependence on the randomly initialized network, especially with a low mask ratio. At lower mask ratios and lower compression rates, the ModNet output also exhibits a more complex distribution, demonstrating the utilization of the random network for representation. 

Additionally, the shallow ModNet layers play a more crucial role, as they are directly connected to the deeper SuperMask layers without rewinding, significantly influencing the synthesis process. In contrast, the deeper ModNet layers provide coarser information, such as lighting and overall structure, which is repeatedly fed into different layers in a rewind fashion. 

\begin{figure*}[t]
    \centering
    \subfloat[]{
        \centering
        \includegraphics[width=0.75\linewidth]{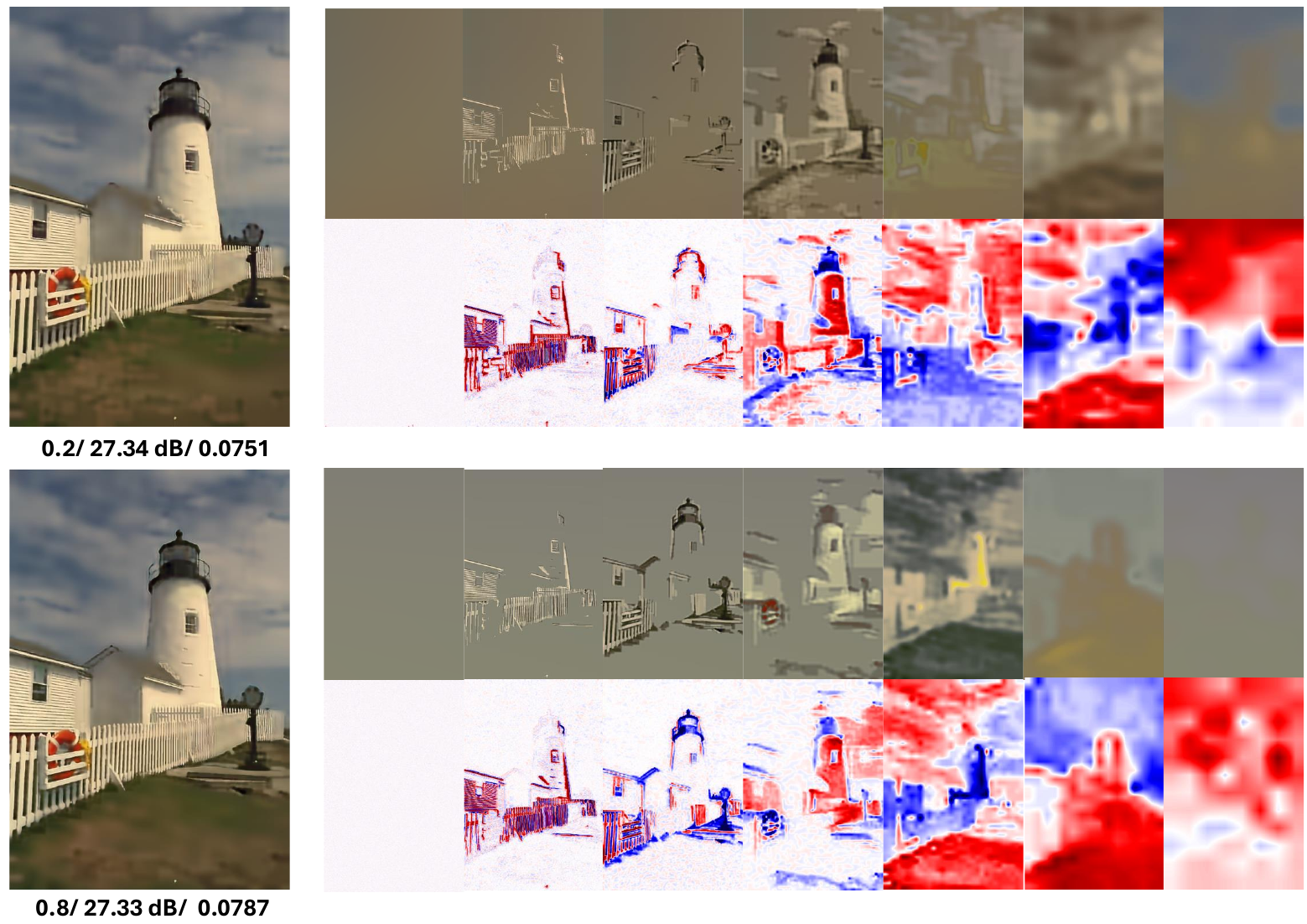}
    }\\
    \subfloat[]{
        \includegraphics[width=0.8\linewidth]{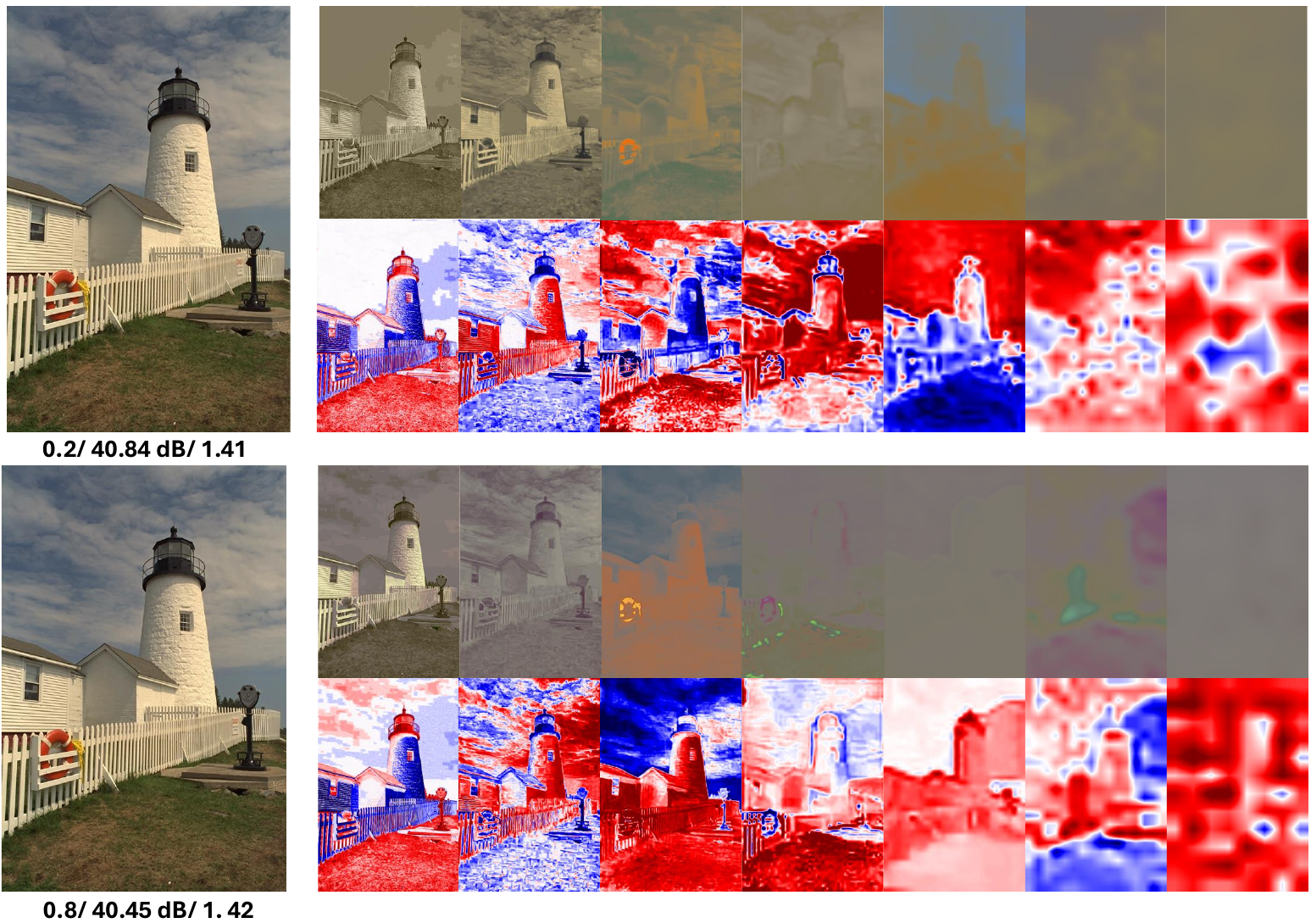}
    }%
    \caption{Visualization of the LotteryCodec scheme using kodim19 as an example, where $0.2/27.34/0.0751$ represent reconstruction with a $0.2$ mask ratio, $27.34$ PSNR, and $0.0751$ bpp. From left to right: reconstructed image followed by latent modulations $\mathbf{z_1}$ to $\mathbf{z_7}$, arranged from high to low resolution. The first row shows reconstructions where all but $\mathbf{z_i}$ latent are set to zero. The second row displays the raw latents, upscaled to match the output resolution. (a). Reconstruction and visualization at a low bit-rate (around 0.078 bpp).  (b). Reconstruction and visualization at a high bit-rate (around 1.41 bpp).}
    \label{vis_all_modulation}
\end{figure*}

\paragraph{Effect of different input latent modulations.} We also visualize the effect of the input latent modulations in Fig. \ref{vis_all_modulation}. As shown in the figure, we observe that the highest resolution latent grid primarily captures luminance and structural details of the image, while lower resolution latent grids provide complementary information for the overall image reconstruction. 

\paragraph{Effect of latent modulations across different bit rates.} Comparing Fig. \ref{vis_all_modulation} (a) and Fig. \ref{vis_all_modulation} (b), we observe that in low-bpp scenarios, lower-resolution modulations play a dominant role in reconstruction, while higher-resolution modulations contribute minimally. This highlights an RD trade-off, where lower-resolution modulations prioritize rate efficiency at the cost of higher distortion. Conversely, in high-bpp scenarios, reconstruction is primarily influenced by higher-resolution modulations, capturing finer image details.

\paragraph{Effect of different mask ratios.} 
Comparing a mask ratio of $0.2$ with $0.8$, we observe that a lower mask ratio distributes more details across each layer of modulation (from $\mathbf{z_1}$ to $\mathbf{z_7}$), allowing the synthesis process to utilize richer structural information. In contrast, at a higher mask ratio of $0.8$, the generation process is predominantly influenced by high-resolution latent representations, with less contribution from lower-resolution modulations.

\paragraph{Interpretation of the design.}
\textcolor{black}{From the above visualizations, we interpret modulation as additional ``bias'', where concatenating neurons with a learned weight mask is equivalent to adding a flexible bias to the subsequent layers. This design demonstrates that the SuperMask network aims to encode images into the network structure, while ModNet introduces a flexible “adaptive bias” to enhance representation flexibility. As shown in Fig. \ref{vis_modulation_effect}, modulation outputs from different ModNet layers contribute varying compensatory information for image reconstruction.}



\nocite{langley00}


\end{document}